\documentclass[unnumsec,webpdf,contemporary,large]{oup-authoring-template}%

\usepackage{adjustbox}

\usepackage{amsmath,amssymb,amsfonts}
\usepackage{longtable}
\usepackage{graphicx}
\usepackage{textcomp}
\usepackage{xcolor}
\usepackage{lipsum}
\usepackage{amsthm}
\usepackage{makecell}
\usepackage{amssymb}
\usepackage{subcaption}
\usepackage{soul}
\usepackage{changes}
\usepackage{amsmath}
\usepackage{amsfonts}
\usepackage{amssymb}
\usepackage[tableposition=top]{caption}
\usepackage[labelfont=bf]{caption}
\DeclareMathOperator*{\argmin}{arg\,min}

\newcommand{\OurAlgo}{QuTIE }

\long\def \ignoreme#1{}

\theoremstyle{thmstyleone}%
\theoremstyle{thmstyletwo}%
\theoremstyle{thmstylethree}%
\newtheorem{definition}{Definition}
\newtheorem{lemma}{Lemma}
\newtheorem{thm}{Theorem}

\begin{document}

\journaltitle{Journal Title Here}
\DOI{DOI HERE}
\copyrightyear{2022}
\pubyear{2019}
\access{Advance Access Publication Date: Day Month Year}
\appnotes{Paper}

\firstpage{1}


\title[]{QuTIE: Quantum optimization for Target
Identification by Enzymes}
\author[1]{Hoang M. Ngo}
\author[1,$\ast$]{My T. Thai}
\author[1,$\ast$]{Tamer Kahveci}

\authormark{}

\address[1]{\orgdiv{Department of Computer and Information Science and Engineering}, \orgname{University of Florida}, \orgaddress{\street{Newell Dr}, \postcode{32611}, \state{Florida}, \country{United States}}}

\corresp[$\ast$]{Corresponding author. \href{email:email-id.com}{email-id.com}}




\abstract{{\em Target Identification by Enzymes (TIE)}
problem aims to identify the set of enzymes in a given metabolic network, such that their inhibition eliminates a given set of target compounds associated with a disease while incurring minimum damage to the rest of the compounds. This is an NP-complete problem, and thus optimal solutions using classical computers fail to scale to large metabolic networks.
In this paper, we develop the first quantum optimization solution, called \emph{QuTIE} (Quantum optimization for Target Identification by Enzymes), to this NP-complete problem. We do that by developing an equivalent formulation of the TIE problem in Quadratic Unconstrained Binary Optimization (QUBO) form. We then map it to a logical graph, and embed the logical graph on a quantum hardware graph.
Our experimental results on 27 metabolic networks
from \emph{Escherichia coli}, \emph{Homo sapiens}, and \emph{Mus musculus} show that \OurAlgo yields solutions which are optimal or almost optimal. Our experiments also demonstrate that \OurAlgo can successfully identify enzyme targets already verified in wet-lab experiments for 14 major disease classes. }



\maketitle

\section{Introduction}\label{sec:intro}


Enzymes catalyze reactions which operate on and transform a set of compounds. The compounds which are input to a reaction are called {\em substrates}, and those that are produced once that reaction completes are called {\em products}. Reactions take place in a complex network topology, where the products of a set of reactions can be substrates for another set of reactions. This organization of reactions is also called {\em metabolic network}. Over- or under-production of certain compounds in metabolism can lead to serious disorders. For instance, the abundance of Dopamine is linked with the development and severity of the  Alzheimer's Disease~\citep{Martorana2014,Pritchard2009}. Similarly, Hunter Syndrome is caused by the metabolism's inability to breakdown sugar molecules~\citep{ijms21041258}, and the malfunction of enzyme \emph{phosphatidic acid phosphatase} leads to overexpression of lipin, causing obesity~\citep{Carman2008}.

One way to address such compound-based disorders is to alter and regulate the production of such compounds by targeting a small subset of enzymes with drugs~\citep{Robertson2007EnzymesAA,Mengqian2020,medicina57111209}, as enzymes are potential drug targets when other drug targets, such as cell surface receptors, DNA, and transporters are not possible to target, or they do not yield the desired impact on the compound regulation~\citep{Robertson2005MechanisticBO}. Drugs which target a specific enzyme inhibit that enzyme and slow down or stop the reactions catalyzed by that enzyme, and thus regulate the abundance of a subset of compounds by stopping/slowing down their production produced downstream of those reactions~\citep{cooney2017,Terentis2010TheSD}. 

Although, it is possible to regulate the metabolic network by targeting enzymes, unintended consequences can happen as a result of this process for various reasons. For instance, the enzymes inhibited by the underlying drug may be responsible from catalysis of multiple reactions. Some of these reactions can produce compounds which lead to the underlying disorder (i.e., intended targets), while others consume/produce different compounds which are unrelated to the disorder (i.e., unintended targets). While reducing the abundance of compounds in the first category is desirable, doing that for the second category may lead to other problems, called {\em side effects}~\citep{Mizutani2012RelatingDI,Sridhar2008}.

One of the fundamental goals in drug development is to obtain a balance between the two potentially conflicting outcomes, namely {\em efficacy} and {\em toxicity} of the drug~\citep{shankarappa2014}. 
Efficacy measures how well the desired outcome (such as lowering the blood pressure if that is the goal of the drug) is achieved, while toxicity measures the damage inflicted on the organism~\citep{Riley2010,Cohenbook2010}.
In order to formulate these concepts mathematically, we call the compounds which are intended to be inhibited (i.e., the compounds whose over-production causes the underlying disorder) \emph{Target Compounds}, and the remaining ones  \emph{Non-target Compounds}. A given enzyme-binding drug limits the production of a set of compounds, some of which are target, while others are not~\citep{Copeland2007TargetingEI}. 
One way to formulate the toxicity of a drug is in terms of the number of non-target compounds which are inhibited (this number is also called \emph{damage})~\citep{Choi2008,Sridhar2008}.

Following from the definitions above, given a metabolic network, including a set of enzymes, reactions, compounds, and relations amongst them, the {\em Target Identification by Enzymes (TIE)} problem aims to identify the set of enzymes, such that their inhibition eliminates a given set of target compounds, while incurring minimum damage. 
This is an NP-complete problem 
~\citep{Song2011} and there are several approaches which address the TIE problem. However, these solutions do not scale well due to exponential complexity of the TIE problem (see Supplementary Materials 1 - SM 1 for details).

Recently, quantum computing has shown its supremacy over classical computers in some tasks which are intractable using classical computers, such as finding prime factors of large integer \citep{Shor1999}. While quantum computing is at a very early stage of development~\citep{Preskill2018}, quantum-inspired methods have already been developed in a wide range of fields such as machine learning \citep{Jerbi2021}, and optimization \citep{Guillaume2022}. 
The fundamental limitation of quantum computing currently is that their capacity in qubits (i.e., quantum bit, which is the quantum analog of the bit classical computers) is limited. One of the most outstanding paradigms which overcomes this limitation for quantum computing is Quantum Annealing (QA). 
This paradigm focuses on solving optimization problems by utilizing quantum fluctuation. QA scales to significantly larger number of qubits than other types of quantum computing. This characteristic enables QA to solve large optimization tasks in bioinformatics such as designing peptides \citep{Mulligan2019}, RNA folding \citep{Dillion2022}, and DNA sequence assembly \citep{Charkiewicz2022}.

QA has three major steps. The first step formulates the optimization problem in \emph{Quadratic Unconstrained Binary Optimization} (QUBO) form. It maps the resulting QUBO on a graph, called \emph{logical graph}, and then maps the logical graph into Quantum Processing Unit (QPU) whose topology is represented by another graph, called \emph{hardware} graph. The final step assigns appropriate parameters to the hardware graph, and runs QA to find candidate solutions for the optimization problem (see the Supplementary Materials 1 - SM 2 for details).

\noindent {\bf Contributions.}
In this paper, we consider the TIE problem, and develop the first quantum optimization solution, called \emph{QuTIE}  (Quantum optimization for Target Identification by Enzymes), to this NP-complete problem. We formulate the TIE problem in QUBO form, and map the enzymes and metabolic reactions to nodes and edges in the logical graph. We utilize QA to find optimal solutions for the TIE problem by mapping the logical graph on the hardware graph. 
We implement and test our solution on the Quantum Hybrid Framework, the largest quantum annealing system available. We compare our method against four methods operating on classic computers: the exact method (OPMET), the IP method, the heuristic method (Double Iterative), and the simulated annealing (SA) method.
Our results on 27 datasets
from \emph{Escherichia coli}, \emph{Homo sapiens}, and \emph{Mus musculus} metabolic network collected from \emph{KEGG} database show that \OurAlgo yields solutions which are optimal, or close to optimal. Our method outperforms the existing methods for large datasets in which the exact method cannot run. Our experiments on the \emph{Biosynthesis of amino
acids} network of Homo sapiens demonstrate that \OurAlgo can successfully identify enzyme targets already verified in wet-lab experiments for 14 major disease classes. 
In addition to solving the NP-complete drug target identification, using quantum optimization, this paper lays the background and opens the door for formulating and solving high complexity problems studying biological networks using quantum computing.




\section{Methods}\label{sec:method}

Here, we first define the TIE problem. We then describe the objective function for the TIE problem in QUBO form. There are four parts in the objective function, namely damage scoring function, target penalty function, reaction inference penalty function, and compound inference penalty function.

\subsection{Formal definition}\label{subsec:def}

Consider a set of enzymes $E$, a set of reactions $R$, and a set of compounds $C$. Metabolic network shows the relationship between the entities in these three sets. Specifically, enzymes catalyze reactions. Each reaction consumes a set of compounds, and produces another set of compounds.
We represent these relationships in a metabolic network as a directed graph $G = (V_G, E_G)$. In this tuple representation, the first term is the union of three mutually exclusive sets of nodes $V_G = E \cup R \cup C$. Each node in $V_G$ corresponds to either an enzyme, reaction, or compound. 
The second term, $E_G$ denotes the set of directed edges among those nodes. Consider two nodes $u$, $v \in V_G$. Each directed edge ($u$, $v$) from node $u$ to $v$ represents one of the three possible types of relations among the nodes as follows:
\begin{enumerate}
    \item The enzyme corresponding to node $u \in E$ catalyses the reaction corresponding to node $v \in R$.
    \item The compound corresponding to node $u \in C$ is a substrate, consumed by the reaction denoted with $v \in R$.
    \item The reaction corresponding to node $u \in R$ produces the compound denoted with $v \in C$.
\end{enumerate}

Notice that, the above mathematical model expresses a metabolic network as a closed system, before the introduction of enzyme binding drug molecules. 
Following from these observations, we list the \emph{inhibition conditions} for each node in $u \in G$ depending on what that node represents as follows:
\begin{itemize}
    \item \textbf{Condition 1:} An enzyme is inhibited when an enzyme binding drug molecule binds to it.
    \item \textbf{Condition 2:} A reaction denoted by node $r \in R$ is inhibited if at least one of the two conditions is satisfied: (i) If there is an input compound, denoted by $c \in C$, consumed by that reaction is inhibited, or (ii) if at least one of the enzymes, denoted by $e \in E$, which catalyzes that reaction is inhibited.
    \item \textbf{Condition 3:} A compound denoted by $c \in C$ is inhibited if all the reactions denoted by $r \in R$ which produce that compound are inhibited.
\end{itemize}

Based on the three conditions above, given a set of compounds, called \emph{target set} $C_{\text{target}} \subseteq C$, it is possible to find a subset of enzymes in $E$ whose inhibition eliminates the production of all compounds in the target set. One can prove the statement above by inhibiting all the enzymes (i.e., all nodes in $E$), thus stopping the production of all the compounds, including those in $C_{\text{target}}$. This however is more than what is needed, as it also eliminates the compounds in $C-C_{\text{target}}$, as well. This is undesirable for the compounds in $C-C_{\text{target}}$ are needed for the healthy metabolism.
We measure the number of compounds in $C-C_{\text{target}}$ whose production stops as a result of inhibition of a subset of enzymes as the \emph{damage} of inhibiting that subset of enzymes (see the example in the Figure~\ref{fig:hypo_img_final} in the Supplementary Materials 1). We desire to inhibit all the compounds in the target set (i.e., maximum efficacy) with minimum damage (i.e., minimum toxicity). Let us denote the set of compounds whose production stops as a result of inhibiting a set of enzymes $E' \subseteq E$ as $C_{E'} \subseteq C$, and the damage to the metabolic network $G$ as the cardinity of the set $C_{E'} - C_{\text{target}}$. We formally define the TIE problem as:

\begin{definition}
Consider a metabolic network consisting of a set of enzymes, a set of reactions, and a set of compounds. Let us denote the set of nodes corresponding to these three sets with $E$, $R$ and $C$ respectively. Let us denote this network with $G = (V_G, E_G)$, where $V_G = E \cup R \cup C$. Given a target set of compounds $C_{\text{target}} \subseteq C$. TIE problem seeks for a set of enzymes $E^\star$ such that:

\begin{equation}
E^\star =  \argmin\{|C_{E'}|:E' \subseteq E \textsf{ AND } C_{\text{target}} \subseteq C_{E'}\}
\end{equation}
\end{definition}\label{def:1}

\subsection{QUBO construction for TIE problem} \label{sec:qubo:form}
Before constructing QUBO for the TIE problem, we develop a Boolean model for the TIE problem. Given the metabolic network $G = (V_G, E_G)$ with $V_G = E \cup R \cup C$, we denote the state of each node $u \in V_G$ in the metabolic network with a binary variable $x_u$ such that:
\begin{equation} \label{eq:indicator:function}
    x_u = \begin{cases}
     0 & \text{if $u$ is not inhibited.} \\
     1 & \text{if $u$ is inhibited.}
\end{cases}
\end{equation}
To satisfy the constraints of the TIE problem, all compounds in the target set $C_{\text{target}}$ need to be inhibited. We express this constraint as:
\begin{equation}\label{eq:sub_add1}
    \prod_{c \in C_{\text{target}}}x_c = 1
\end{equation}
Next, we present inhibition conditions as follows:
\begin{itemize}
    \item \textbf{Condition 1:} The state of an enzyme $e \in E$ only depends on $x_e$.
    \item \textbf{Condition 2:} Consider a node in the metabolic network corresponding to a reaction, $r \in R$. Let us define the set of nodes in its immediate upstream with the set $N(r)$. Recall that a node $v \in N(r)$ if one of the two criteria is satisfied: (i) $v$ corresponds to an enzyme which catalyzes the reaction corresponding to node $r$, and (ii) $v$ corresponds to a compound which is consumed by the reaction corresponding to node $r$. Thus, we have $v \in (E \cup C)$.  If any of the nodes in $N(r)$ are inhibited, that implies $r$ is also inhibited. As a result, the state of the reaction $r$ is valid if it satisfies:
    \begin{equation}\label{eq:sub1}
    x_r = 1 - \prod_{v \in N(r)}(1-x_{v})
    \end{equation}
    \item \textbf{Condition 3:} Consider a node in the metabolic network corresponding to a compound, $c \in C$. Let us define the set of nodes in its immediate upstream with set $N(c)$. Recall that each node $v \in N(c)$ corresponds to a reaction that produces the compound denoted by node $c$, thus $v \in R$. If all of the nodes in $N(c)$ are inhibited, then $c$ is also inhibited. As a result, a state of a compound $c$ is valid if it satisfies:
    \begin{equation}\label{eq:sub6}
        x_c = \prod_{v \in N(c)} x_{v}
    \end{equation}
\end{itemize}
We acknowledge the presence of a scenario where a group of alternative enzymes catalyzes a reaction, and the inhibition of the reaction occurs only when all enzymes in the group are inhibited. In such cases, we can reformulate Equation~(\ref{eq:sub1}) in a similar manner to Equation~(\ref{eq:sub6}).

We consider an assignment of values to the set of variables $\{x_u \mid u \in E \cup R \cup C\}$ is valid if all $x_u$ satisfy~Constraints (\ref{eq:sub_add1}), (\ref{eq:sub1}), and (\ref{eq:sub6}).

The critical challenge we need to address to solve the TIE problem by QA is to represent the TIE problem as a QUBO function. Thus, our goal is to design an energy function $H$ for the TIE problem in form of QUBO. $H$ takes a set of binary variables as input (the input binary set also includes auxiliary binary variables which we discuss later). As $H$ is unconstrained, we need to discriminate invalid and valid assignments of values to the input variables of this function. Furthermore, the function $H$ must return values corresponding to the damage of input assignment. To sum up, in order to model the TIE problem, function $H$ must follow two principles:
\begin{itemize}
    \item The value of $H$ for a valid assignment must be lower than that for every invalid assignments.
    \item The value of $H$ for a valid assignment must be equal to the damage produced by that assignment.
\end{itemize}
Minimizing the value of function $H$ which follows these two principles is equivalent to finding a valid assignment with minimum damage for the underlying TIE problem.

Based on two above principles, we construct the energy function $H$ as a combination of four quadratic functions, named \emph{Damage Scoring}, \emph{Target Penalty}, \emph{Reaction Inference Penalty}, and \emph{Compound Inference Penalty}. Damage Scoring function measures the damage of the input assignment. Target Penalty function ensures that all target compounds are inhibited (see Constraint~(\ref{eq:sub_add1})). Reaction Inference Penalty function controls the inhibition condition of reactions (see Constraint~(\ref{eq:sub1})). Compound Inference Penalty function ensures the inhibition condition of compounds (see Constraint~(\ref{eq:sub6})). Next, we elaborate on construction of these functions, and explain how they fit to corresponding constraints.


\noindent{\bf Damage scoring function.}
This function models the toxicity arising from the disturbance of the production of those compounds which are not intended to be inhibited, but are inhibited as a result of inhibiting a subset of enzymes in the given network $G$. We compute the damage scoring function in term of the binary variables $x_c$ ($c \in C$) with a positive constant $k_1$ as (see Lemma \ref{lemma:damage} in Supplementary Materials 1 - SM 4):
\begin{equation}\label{eq:main1}
    H_{\text{damage}} = k_1\sum_{c \in C-C_{\text{target}}} x_c
\end{equation}


\noindent{\bf Target penalty function.}
This function models the loss in the efficacy of the drug by representing the constraint that requires inhibition of all the compounds in the target set (see constraint~(\ref{eq:sub_add1})). We write this function in term of the binary variables $x_c$ above with a positive constant $k_2$ as follows:

\begin{equation}\label{eq:main2}
    H_{\text{target}} = k_2\sum_{c \in C_{\text{target}}} (1 - x_c)
\end{equation}

The target penalty $H_{\text{target}}$ in Equation~(\ref{eq:main2}) is minimized if and only if the states of all compounds in the target set are 1 (i.e., when all the targeted compounds are inhibited). In other words, the function $H_{\text{target}}$ only returns minimum value for valid assignment of $\mathbf{x}$ (see Lemma \ref{lemma:target} in Supplementary Materials 1 - SM 4).

\noindent{\bf Reaction inference penalty.}
This function expresses the second constraint in the inhibition conditions (see Equation~(\ref{eq:sub1})). Given a reaction $r \in R$, we construct a system of two linear inequalities from Equation~(\ref{eq:sub1}) as:

\begin{equation}\label{eq:sub2}
    x_r \geq x_{v} \forall v \in N(r)
\end{equation}
\begin{equation}\label{eq:sub3}
    0 \geq x_r - \sum_{v \in N(r)} x_{v}
\end{equation}
Since each variable $x_r$ above takes value either 0 or 1, in order to satisfy Inequality~(\ref{eq:sub2}), $\forall v \in N(r)$, we must have $x_r - x_{v} = 0$, or $x_r - x_{v} - 1 = 0$. Following from these two observations, we construct a quadratic expression for Inequality~(\ref{eq:sub2}) as follows:
\begin{equation}\label{eq:sub4}
    \sum_{v \in N(r)}[(x_r - x_{v})^2 + (x_r - x_{v} - 1)^2] - |N(r)|
\end{equation}
Expression~(\ref{eq:sub4}) obtains the minimum value of 0 if and only if $x_u$, and $x_{v}$ with $ v \in N(u)$ satisfy Inequality~(\ref{eq:sub2}).

Building a quadratic equation for modeling  Inequality (\ref{eq:sub3}) is nontrivial. This is because the value of difference ($x_r - \sum_{v \in N(r)} x_{v}$) depends on the number of nodes in $N(r)$. To overcome this challenge, we define $|N(r)| + 1$ auxiliary binary variables $t_{r0}$, $t_{r1}$, \dots, and write the following expression.
\begin{equation}\label{eq:sub5}
    \Biggr[x_r - \sum_{v \in N(r)}x_{v} + \sum_{\alpha=0}^{|N(r)|}(\alpha t_{r\alpha}) \Biggr]^2 + (1-\sum_{\alpha=0}^{|N(r)|}t_{r\alpha})^2
\end{equation}
In Expression~(\ref{eq:sub5}), each of the $|N(r)| + 1$ auxiliary binary variables $t_{r\alpha}$ models one of the $|N(r)| + 1$ possible valid values the right hand side of Inequality~(\ref{eq:sub3}) can take. Therefore, Expression~(\ref{eq:sub5}) reaches to value of 0 if and only if states $x_r$ and $x_{v}$, with $v \in N(r)$ satisfy Inequality~(\ref{eq:sub3}), and the variable $t_{r\alpha} = 1$ only if $x_r - \sum_{v \in N(r)} x_{v} + \alpha = 0$.

Using Expressions (\ref{eq:sub4}) and (\ref{eq:sub5}) for reaction $r \in R$, we construct the reaction inference penalty function which ensures the second constraint from the inhibition conditions using a tunable positive constant $k_3$, and constant $\lambda_R$ as: 



\begin{multline} \label{eq:main3}
    H_{\text{reaction}} = k_3\sum_{r \in R}\Biggl\{\sum_{v \in N(r)}[(x_r - x_{v})^2 + (x_r - x_{v} - 1)^2] \\ 
    + \Biggl[x_r - \sum_{v \in N(r)}x_{v} + \sum_{\alpha=0}^{|N(r)|}(\alpha t_{r\alpha}) \Biggr]^2 + (1-\sum_{\alpha=0}^{|N(r)|}t_{r\alpha})^2 \Biggr\} + \lambda_R
\end{multline}

The function $H_{\text{reaction}}$ returns minimum value of 0 only for valid assignment of $\mathbf{x}$, and auxiliary variable $t$ (see Lemma \ref{lemma:reaction} in Supplementary Materials 1 - SM 4).

\noindent{\bf Compound inference penalty.}
This function models the third, and the final constraint in the inhibition conditions (see the Equation~(\ref{eq:sub6})). Given a compound $c \in C$, we construct a system of two linear inequalities from Equation~(\ref{eq:sub6}) as:
\begin{equation}\label{eq:sub7}
    x_c \leq x_{v}\forall v \in N(c)
\end{equation}
\begin{equation}\label{eq:sub8}
    -|N(c)| \leq x_c - \sum_{v \in N(c)}x_{v} - 1
\end{equation}
In order to satisfy Inequality~(\ref{eq:sub7}), $\forall v \in N(c)$, we must have $x_c - x_{v} = 0$, or $x_c - x_{v} + 1 = 0$. These two observations lead to a the following quadratic expression:

\begin{equation}\label{eq:sub9}
    \sum_{v \in N(c)}[(x_c - x_{v})^2 + (x_c - x_{v} + 1)^2] - |N(c)|
\end{equation}
Expression~(\ref{eq:sub9}) yields minimum value of 0 if and only if $x_c$ and $\forall v \in N(c)$,  $x_{v}$ satisfy Inequality~(\ref{eq:sub7}). The proof of this statement is the same as that for Equation~(\ref{eq:sub4}).

To build a quadratic expression for Inequality~(\ref{eq:sub8}), similar to the Expression~(\ref{eq:sub5}), we define $|N(c)|+1$ auxiliary binary variables $w_{c0}$, $w_{c1}$, \dots. We have the quadratic expression for Inequality~(\ref{eq:sub8}) as follows:

\begin{equation}\label{eq:sub10}
    \Biggl[ x_c - \sum_{v \in N(c)}x_{v}  - 1 + \sum_{\beta=0}^{|N(c)|}(\beta w_{c\beta}) \Biggr]^2 + (1-\sum_{\beta=0}^{|N(c)|}w_{c\beta})^2
\end{equation}
We observe from Expression~\ref{eq:sub10} that $|N(c)|+1$ auxiliary binary variables $w_{c\beta}$ correspond to $|N(c)|+1$ possible valid values the right hand side of Inequality~(\ref{eq:sub8}) can take. Therefore, Expression~(\ref{eq:sub10}) takes minimum value of 0 if and only if states $x_c$ and $x_{v}$ satisfy Inequality~(\ref{eq:sub8}), and the variable $w_{c\beta} = 1$ only if $x_c - \sum_{v \in N(c)} x_{v} - 1 + \beta = 0$.

Using Expressions (\ref{eq:sub9}) and (\ref{eq:sub10}) for compound $c \in C$, we construct the compound inference penalty with the help of a tunable positive constant $k_4$, and constant $\lambda_C$ as: 
\begin{multline} \label{eq:main4}
    H_{\text{compound}} = k_4\sum_{c \in C} \Biggl\{ \sum_{v \in N(c)}[(x_c - x_{v})^2 + (x_c - x_{v} + 1)^2] \\
    + \Biggl[ x_c - \sum_{v \in N(c)}x_{v} - 1 + \sum_{\beta=0}^{|N(c)|}(\beta w_{c\beta})\Biggr]^2 + (1-\sum_{\beta=0}^{|N(c)|}w_{c\beta})^2\Biggr\} + \lambda_C
\end{multline}
The function $H_{\text{compound}}$ returns minimum value of 0 only for valid assignment of $\mathbf{x}$, and auxiliary variable $w$ (see Lemma \ref{lemma:compound} in Supplementary Materials 1 - SM 4).

Combining function (\ref{eq:main1}), (\ref{eq:main2}), (\ref{eq:main3}), and (\ref{eq:main4}), we represent the TIE problem in QUBO form by an energy function as follows:
\begin{equation}
\label{eq:objective}
    H = H_{\text{damage}} + H_{\text{target}} + H_{\text{reaction}} + H_{\text{compound}}
\end{equation}

We observe that the function $H$ is a combination of linear, and quadratic forms. Because binary variable $x$ satisfies $x = x^2$, we rewrite the linear term $x$ as $x^2$ making the entire of equation quadratic. As a result, the function $H$ is in QUBO form which can be processed by QA. The final result we expect to obtain after QA process is a set of variables $Y^{\star} = X \cup \{t_{r\alpha}|r \in R, \alpha \in N(r)\} \cup \{w_{c\beta}|c \in C, \beta \in N(c)\}$ such that:
\begin{equation} \label{eq:optvalue}
    Y^{\star} = \argmin \{H\}
\end{equation}

The function $H$ models the TIE problem because it satisfies the two principles we mentioned before in this section. Penalty functions $H_{\text{target}}$, $H_{\text{reaction}}$, and $H_{\text{compound}}$ return the minimum value of 0 for only valid assignments. As a result, if we choose positive constants $k_2, k_3, k_4$ which are large enough, the outputs of the function $H$ for valid assignments are always lower than those for invalid assignments. In addition, the function $H_{\text{damage}}$ returns the damage corresponding to the input assignment if we set $k_1 = 1$. Thus, for valid assignments whose penalty scores are always equal to 0, the function $H$ returns corresponding damage of those valid assignments (see Theorem \ref{theorem:objective} in Supplementary Materials 1 - SM 4).

\section{Discussion}\label{sec:exp}


In this section, we evaluate our method on a small dataset and a large dataset. We describe datasets in details in the Supplementary Materials 1 - SM 3. We compare our method to four methods: 
\begin{itemize}
    \item \textbf{Exact method (OPMET):} This method uses branch-and-bound to examine all possible combinations of inhibited enzymes, and thus it is optimal~\citep{Sridhar2008}.
    \item \textbf{Integer Programming (IP)}: We use the IP formulation for the \emph{BN-ReactionCut} problem~\citep{Tamura2011} with a modification in its objective function to solve TIE problem. \emph{BN-ReactionCut} problem makes a simplifying, but incorrect assumption that each reaction is controlled by one enzyme. As a result, it ignores the set of enzymes operating on the metabolic network. In order to make fair comparisons with our method, we post-process the solution returned by IP, and randomly select one enzyme corresponding to each inhibited reaction from the resulting solution for inhibition. We then calculate actual damage caused by inhibiting selected enzymes. For each test case, we perform process for a constant number of times, and report the average damage.
    \item \textbf{Heuristic (Double Iterative)}: This is a heuristic variant of OPMET working in two phases~\citep{Song2009}.
    \item \textbf{Simulated Annealing (SA)}:  This method is  inspired by the annealing process, similar to QA. However, unlike QA, SA works on a classical computer. We run SA with the same objective function $H$ that we use in QA.
\end{itemize}

\subsection{Evaluation using synthetically selected targets}

Our first set of experiments answer the question: How does \OurAlgo perform under different network characteristics and number of target compounds compared to existing solutions? 

\noindent{\bf Experimental setup for Quantum Hybrid Solver.} We use a \emph{Quantum Hybrid Solver} provided by D-Wave to solve our proposed QUBO. It is a hybrid framework combining classical and quantum computing techniques to find optimal solutions for a given QUBO formulation. We explain this framework in details in Supplementary Materials 1 - SM 2. We set the running time limit of the Quantum Hybrid Solver to 10 minutes for the small datasets, and 20 minutes for the large ones. In all experiments, solutions provided by Quantum Hybrid Solver are valid (i.e., all target compounds are eliminated).

\noindent{\bf Experimental setup for target selection.} Let us denote the number of target compounds to be inhibited with $k$. Given a metabolic network, we run experiments by growing the number of target compounds to be inhibited in that network
from $k =$ 2 to 27, at increments of 5 (i.e., six different values of $k$) by randomly selecting $k$ target compounds from that network. We repeat this procedure up to 5 times for each combination of metabolic network and target network size, measure damage and running time, and report the average.

\noindent{\bf Experimental setup for datasets.} We use metabolic pathways for three species: Escherichia coli (eco or E.Coli), Homo sapiens (hsa or H.Sapiens), and Mus musculus (mmu or M.Musculus) from the KEGG database~\mbox{\citep{Kanehisa2000}}. We categorize these metabolic networks into two groups based on the number of interactions in each: small and large pathways. The number of nodes in small pathways ranges from 35 to 93, while the number of nodes in large pathways ranges from 146 to 305. Tables ~\mbox{\ref{table:data_small}} and ~\mbox{\ref{table:data_large}} in Supplementary Materials 1 - SM 2 list the characteristics of pathways in more details.

\begin{figure}[t]
\begin{subfigure}{.2\textwidth}
  \centering
  \includegraphics[width=1\textwidth]{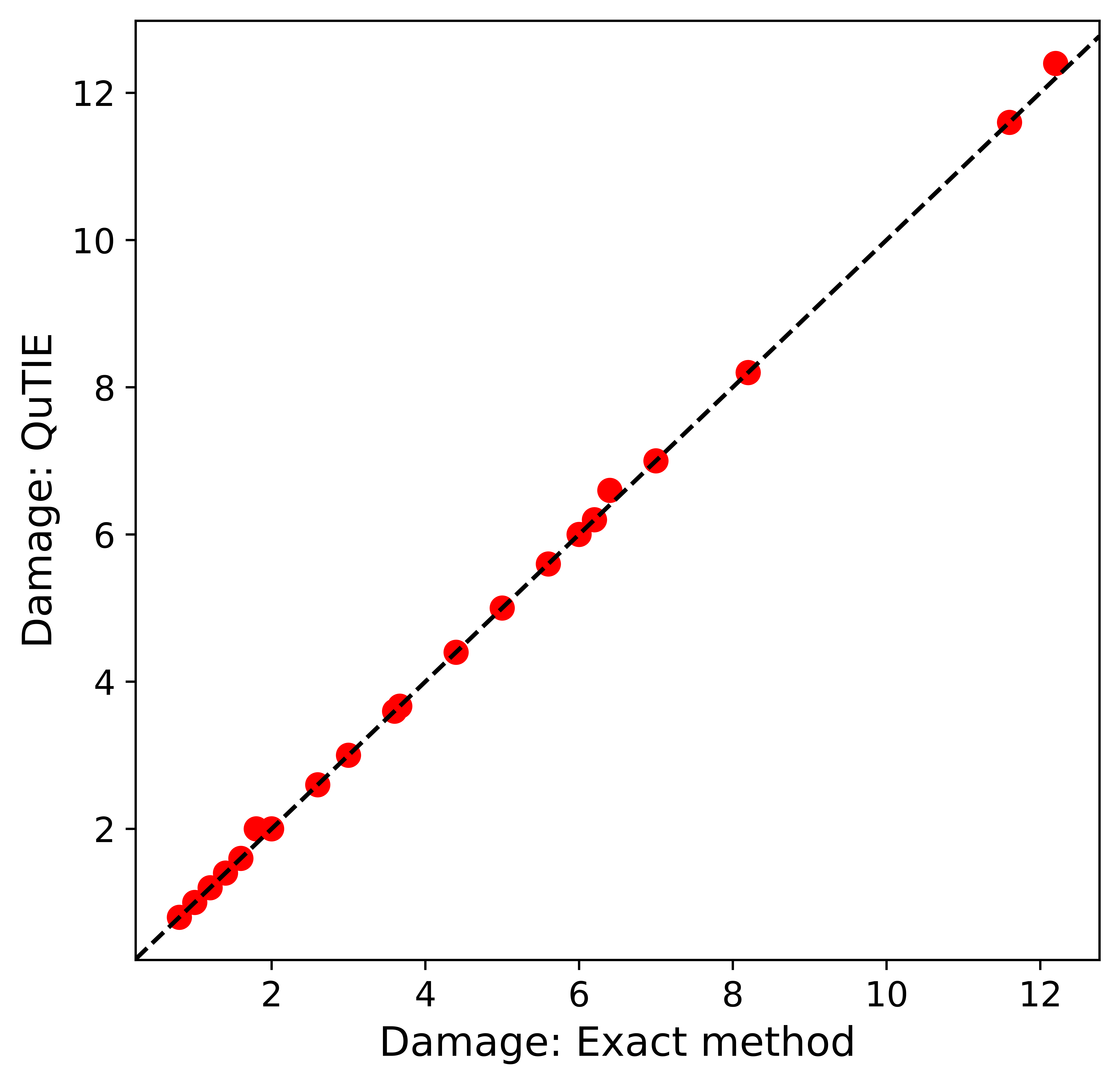}
  \caption{\emph{E.Coli}}
  \label{fig:sfig1:eco}
\end{subfigure}
\hfill
\begin{subfigure}{.2\textwidth}
  \centering
  \includegraphics[width=1\columnwidth]{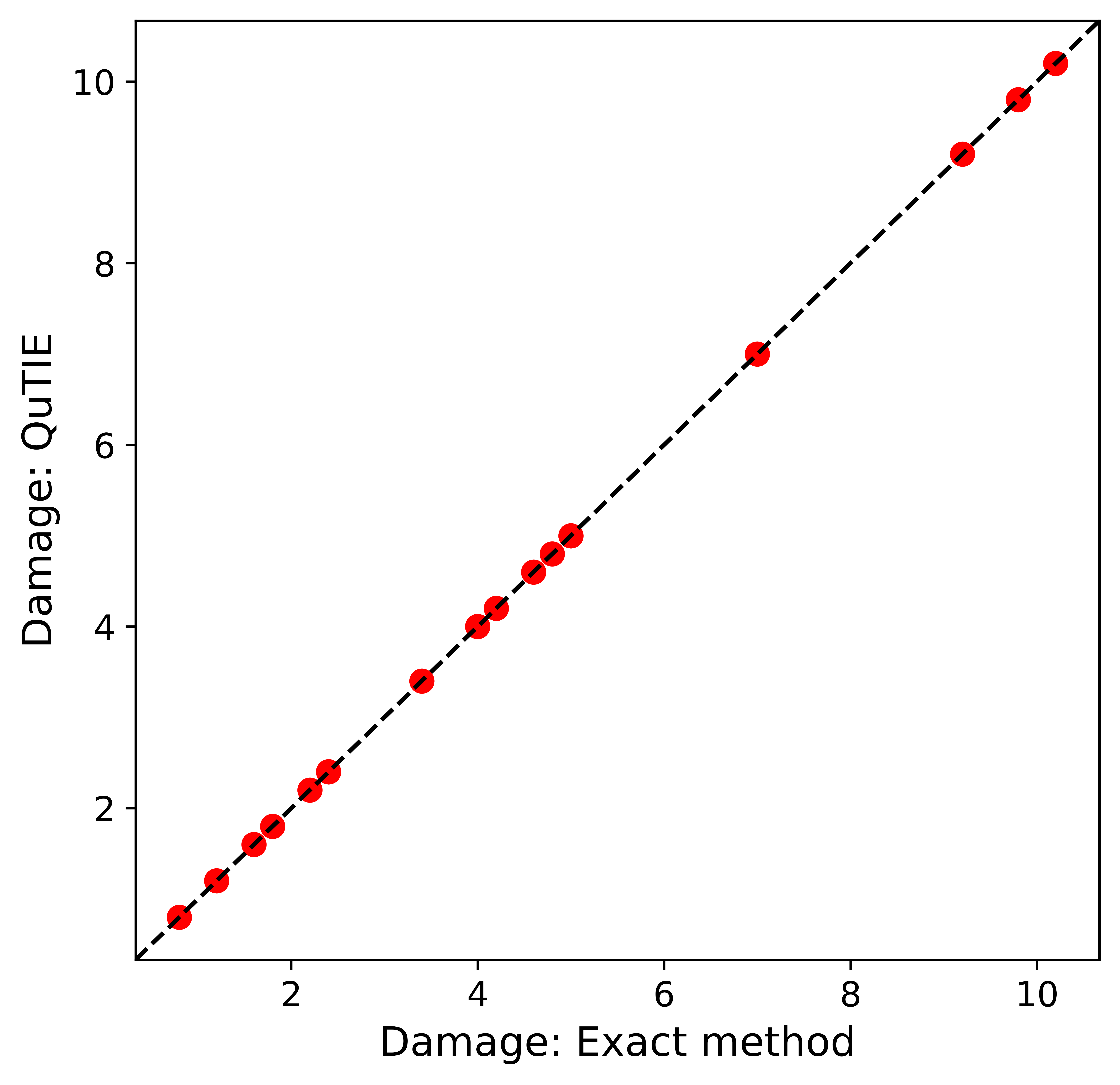}
  \caption{\emph{H.Sapiens}}
  \label{fig:sfig1:hsa}
\end{subfigure}
\hfill
\begin{subfigure}{.2\textwidth}
  \centering
  \includegraphics[width=1\columnwidth]{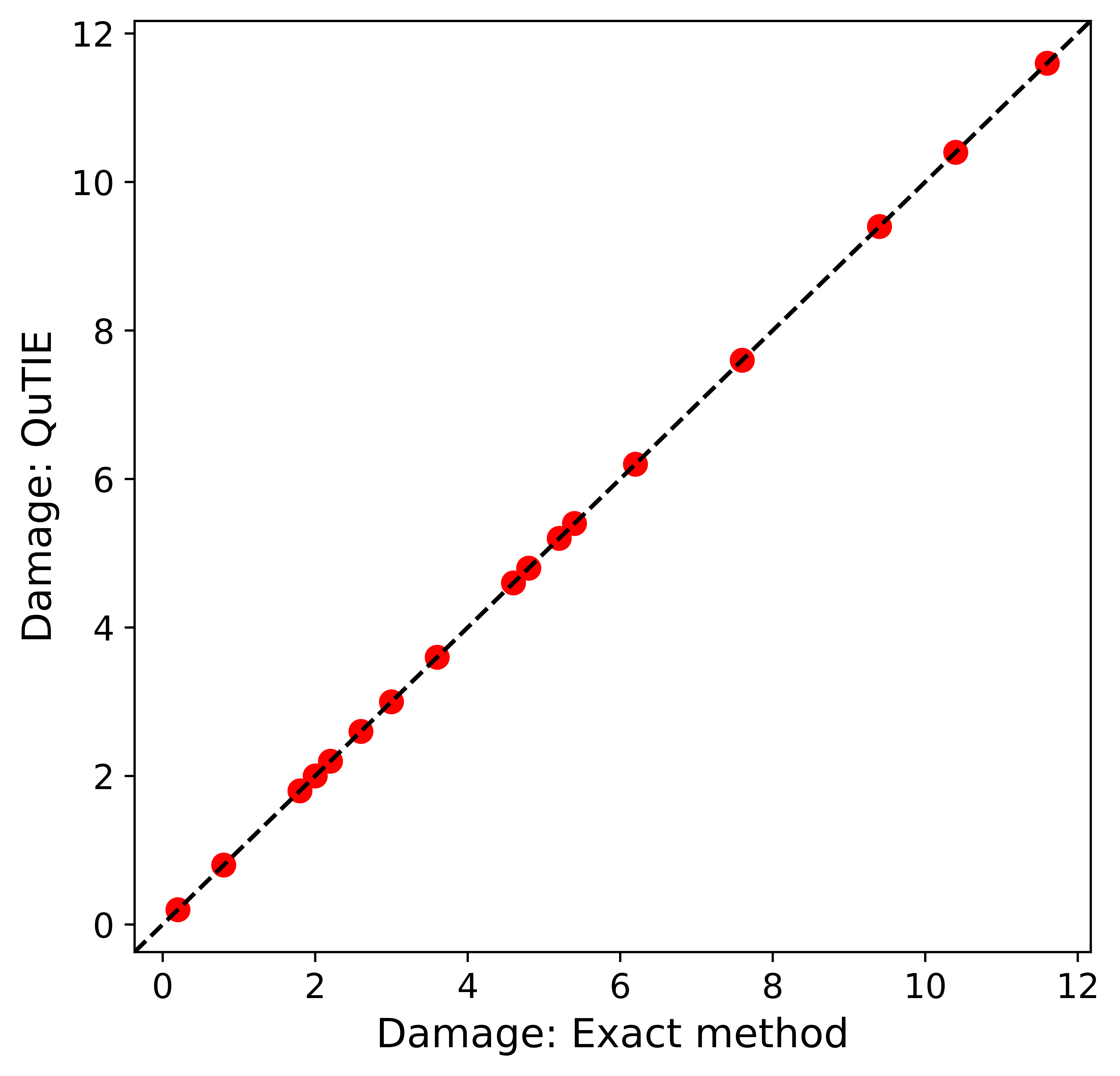}
  \caption{{\em M.Musculus}}
  \label{fig:sfig1:mmu}
\end{subfigure}
\hfill
\begin{subfigure}{.2\textwidth}
  \centering
\hspace*{-24pt}  \includegraphics[width=1\columnwidth]{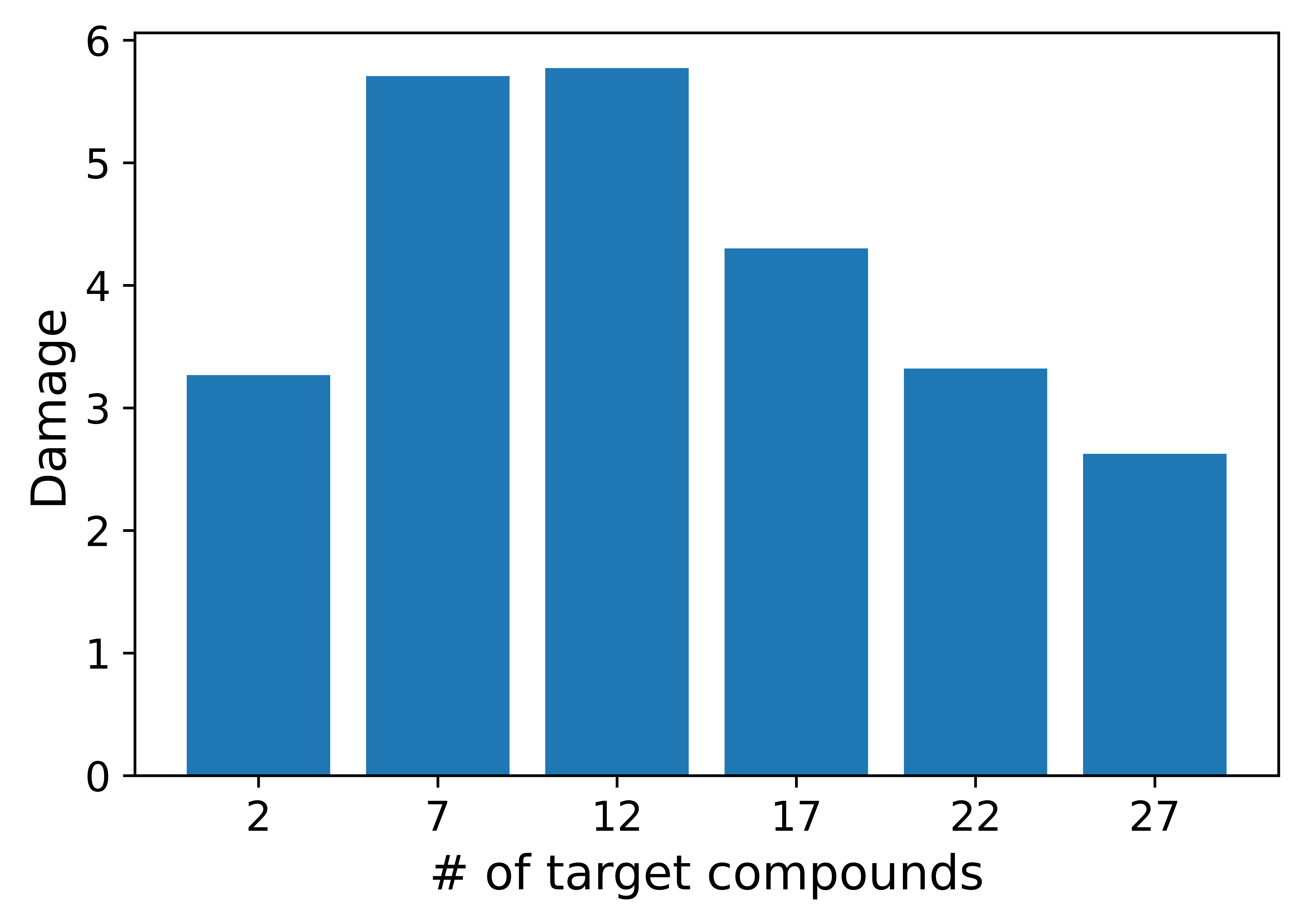}
  \caption{}
  \label{fig:sfig1:k}
\end{subfigure}
\hfill
\begin{subfigure}{.2\textwidth}
  \centering
  \includegraphics[width=1\columnwidth]{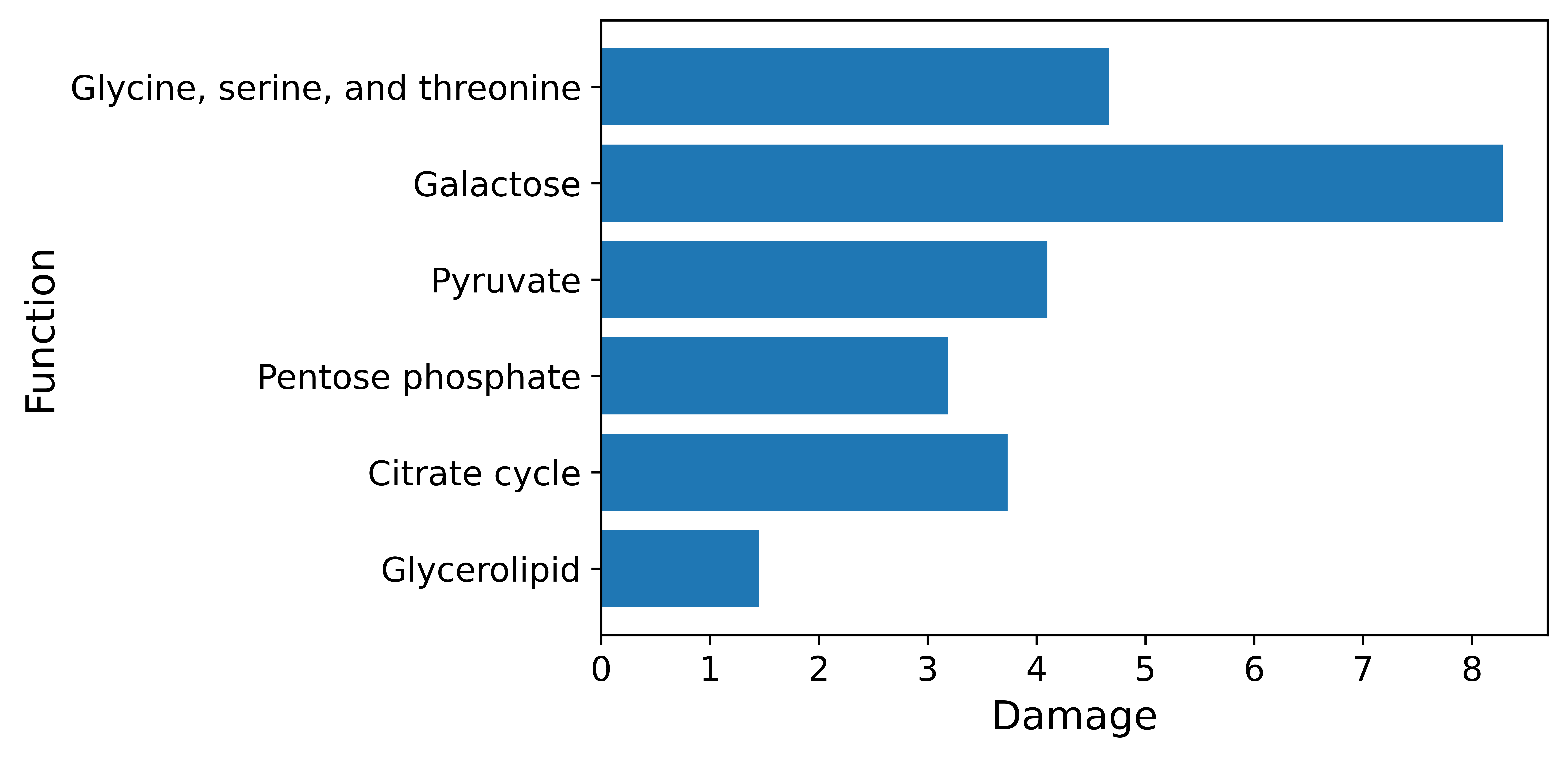}
  \caption{}
  \label{fig:sfig1:function}
\end{subfigure}
\hfill
\begin{subfigure}{.2\textwidth}
  \centering
  \includegraphics[width=1\columnwidth]{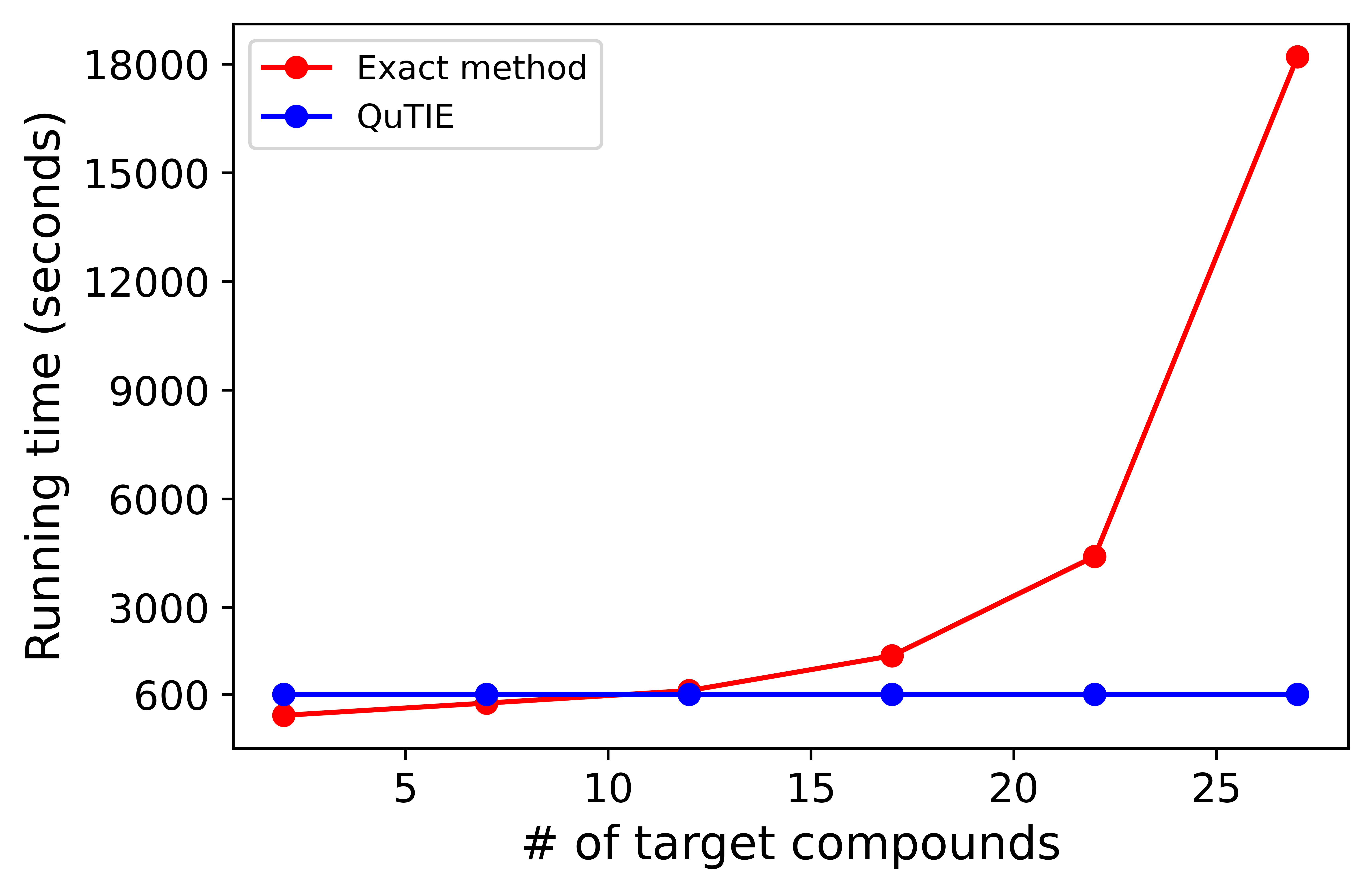}
  \caption{}
  \label{fig:sfig1:time}
\end{subfigure}

\caption{Analysis of \OurAlgo on small datasets. (a),(b),(c) Damage values provided by \OurAlgo and the exact method for the three species. Each point corresponds to the average of a combination of one network and one $k$ value across all test cases. The diagonal line is the $x = y$ line. (d) Average damage value of \OurAlgo across all parameters grouped by the number of target compounds. (e) Average damage value of \OurAlgo across all parameters grouped by the network function.
(f) Comparison between \OurAlgo and the exact method on small datasets in term of running time. 
} 

\end{figure}

\subsubsection{Comparison with the exact method} \label{sec:expt:exact}
Here, we examine the performance of \OurAlgo on small datasets by comparing to the exact method, OPMET. In the small datasets, we do not include the \emph{Pyruvate} metabolic network of \emph{E.Coli}, and the \emph{Glycine, serine, and threonine} metabolic networks of \emph{H.Sapiens}, and \emph{M.Musculus} because their sizes are too big for the exact method to run. Notice that the exact method guarantees optimal solutions. Thus, the fundamental purpose of this comparison is to observe (1) how well \OurAlgo optimizes the damage function for the TIE problem under different metabolic networks in the small datasets, and various target compound set sizes, with respect to the optimal solution, and (2) how much  running time  \OurAlgo needs to arrive at this solution as compared to the exact method. We use the small networks listed in Table~\ref{table:data_small} for the exact method does not scale to larger networks. In total, we perform 900 experiments (i.e., 5 networks $\times$ 3 species $\times$ 6 values of $k$ $\times$ 5 random repetitions $\times$ 2 methods).

Recall that the TIE problem aims to minimize the damage, while inhibiting all target compounds. We, first compare the two methods in term of the damage their results inflict on the given metabolic network. We obtain the average damage of solutions from \OurAlgo and the exact method for each combination of metabolic networks and target compound set size. Figures \ref{fig:sfig1:eco}, \ref{fig:sfig1:hsa}, and \ref{fig:sfig1:mmu} present the results for three species including E.Coli, H.Sapiens, and M.Musculus respectively. Each point in this figure corresponds to the average damage of one (network, target compound set size) pair. Our results demonstrate that \OurAlgo is able to obtain the optimal or near optimal solutions for all experimental settings. Figure \ref{fig:sfig1:function} illustrates the average damage of experimental settings for the same network function. From the results, we observe that inhibiting target compounds from \emph{Galactose} metabolic networks can cause more damage than those from any other functions. In addition, we examine the average damage for different number of target compounds, and show the results in Figure~\ref{fig:sfig1:k}. We observe a upward trend in the average damage when the number of target compounds is small. We infer that inhibiting medium-sized sets of target compounds may cause the most damage to the network.

One of the fundamental promises of quantum optimization algorithms is that they can solve problems with high complexity dramatically faster than the algorithms operating on traditional computers. Following from this, and our observation above that \OurAlgo yields optimal damage values even for very large values of $k$, the next important question we need to answer is at what running time cost does our algorithm achieve these results, for the TIE problem is NP-complete?

We compare the running time of the two methods. The running time of \OurAlgo is the time limit set in the Quantum Hybrid Solver (10 minutes) in all experiments for the small dataset. For the exact solution, we report the average running time for each size of target compound set. We report the running time of two methods in Figure~\ref{fig:sfig1:time}. Our results suggest that the number of target compounds $k$ has massive impact on the total running time of the exact method on classical computers. This is expected as the complexity of the TIE problem is exponential in the target compound set size in the worst case. For small $k$, finding optimal solutions is trivial. As the value of $k$ increases though, the running time to find exact solutions quickly becomes impractical. The second observation is that \OurAlgo on the other hand, is not affected by the value of $k$, and it yields optimal solutions in the preset time limit. Although we cannot claim anything about precise quantum speed-up over classical exact method due to fixed time limit, quantum computing shows its potential power for reaching optimal solutions in a fast manner. To be clear, in the case with $k = 27$, \OurAlgo can find optimal solutions in a duration which is only equal to nearly 3\% of the running time of the exact method. The gap between the running time of our method and that of the exact solution grows with increasing value of $k$. 

\begin{figure}[t]
\begin{subfigure}{.15\textwidth}
  \centering
  \includegraphics[width=1\textwidth]{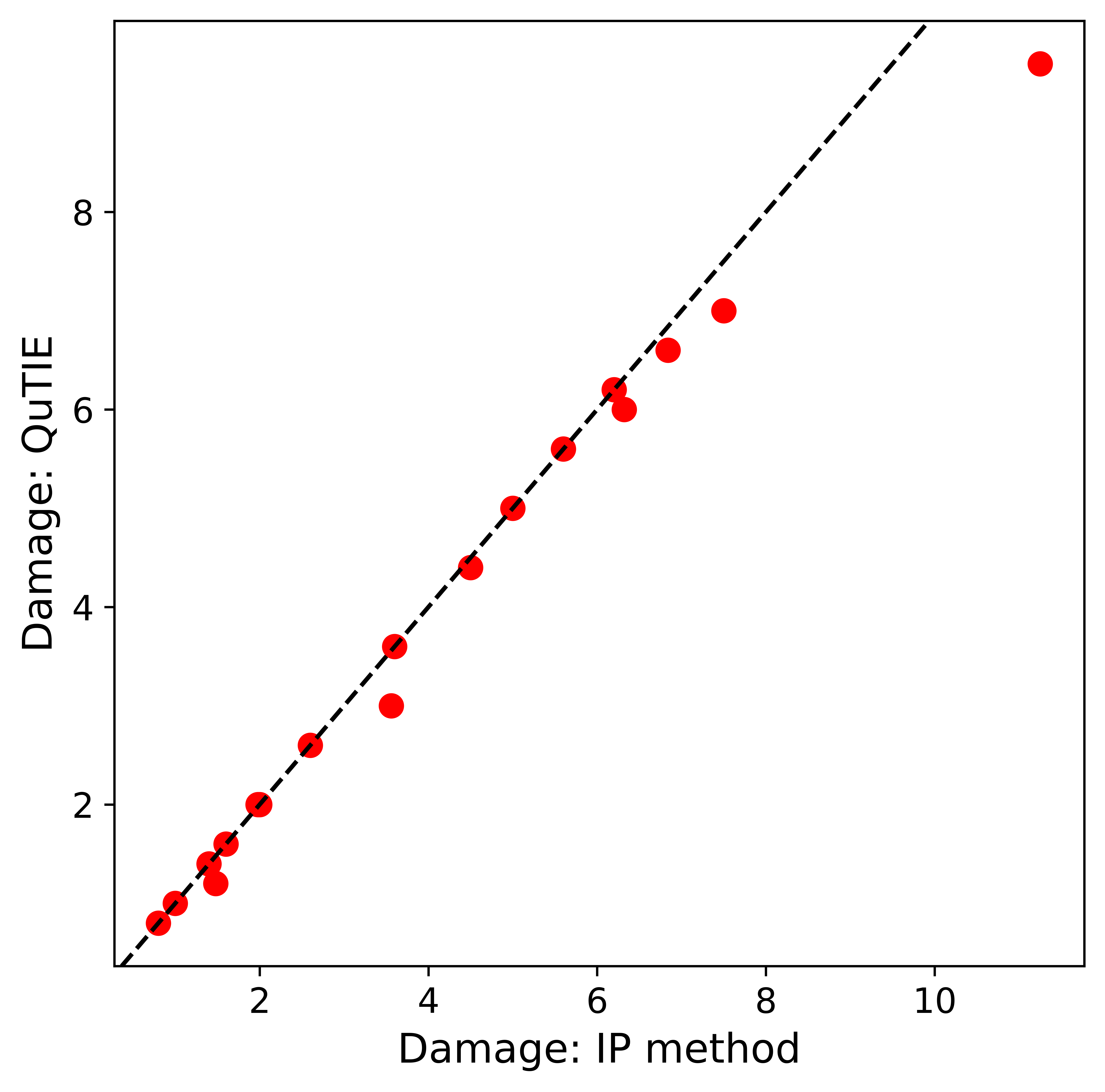}
  \caption{\emph{E.Coli}}
  \label{fig:sfig1_add:eco}
\end{subfigure}
\hfill
\begin{subfigure}{.15\textwidth}
  \centering
  \includegraphics[width=1\columnwidth]{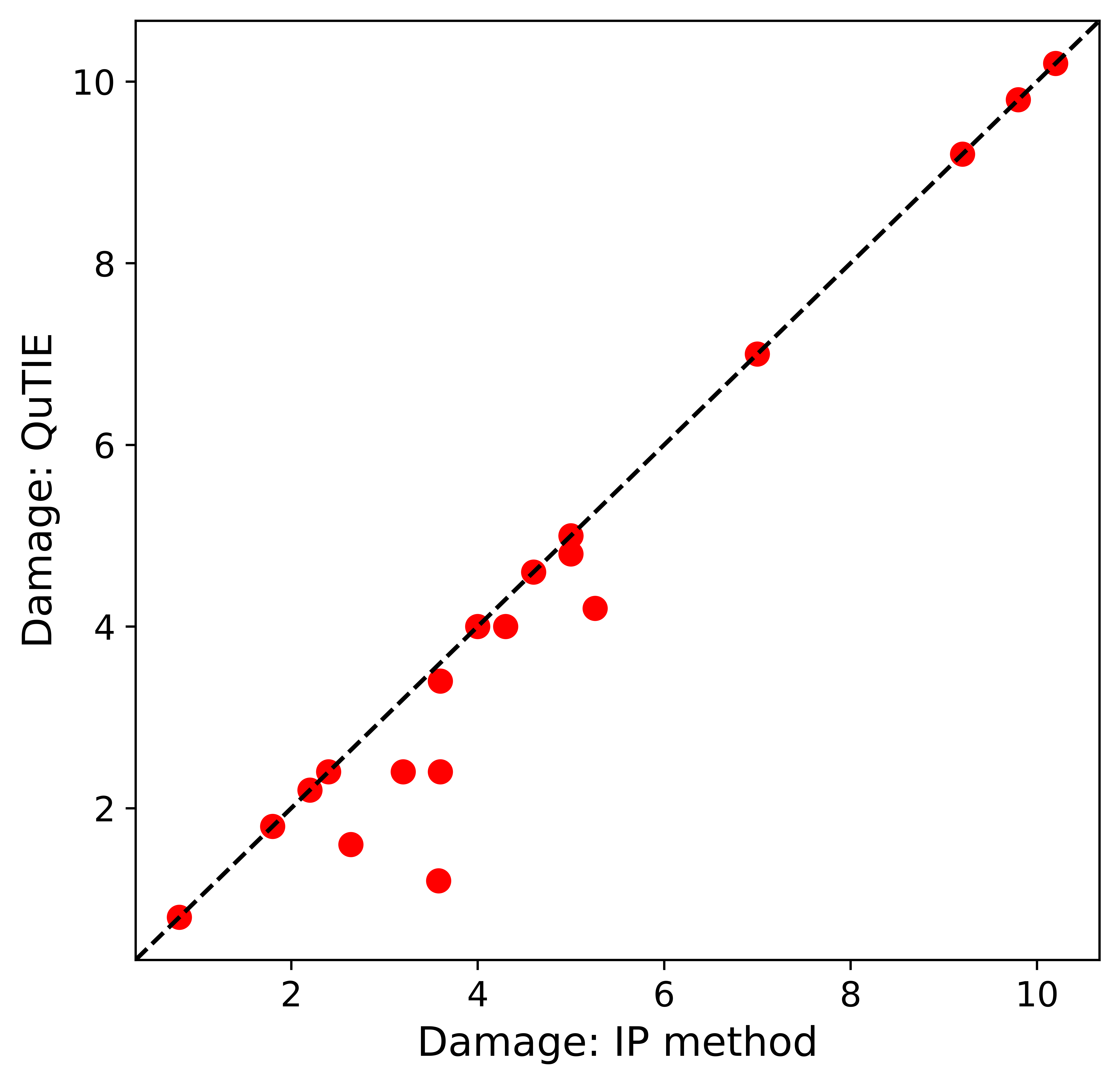}
  \caption{\emph{H.Sapiens}}
  \label{fig:sfig1_add:hsa}
\end{subfigure}
\hfill
\begin{subfigure}{.15\textwidth}
  \centering
  \includegraphics[width=1\columnwidth]{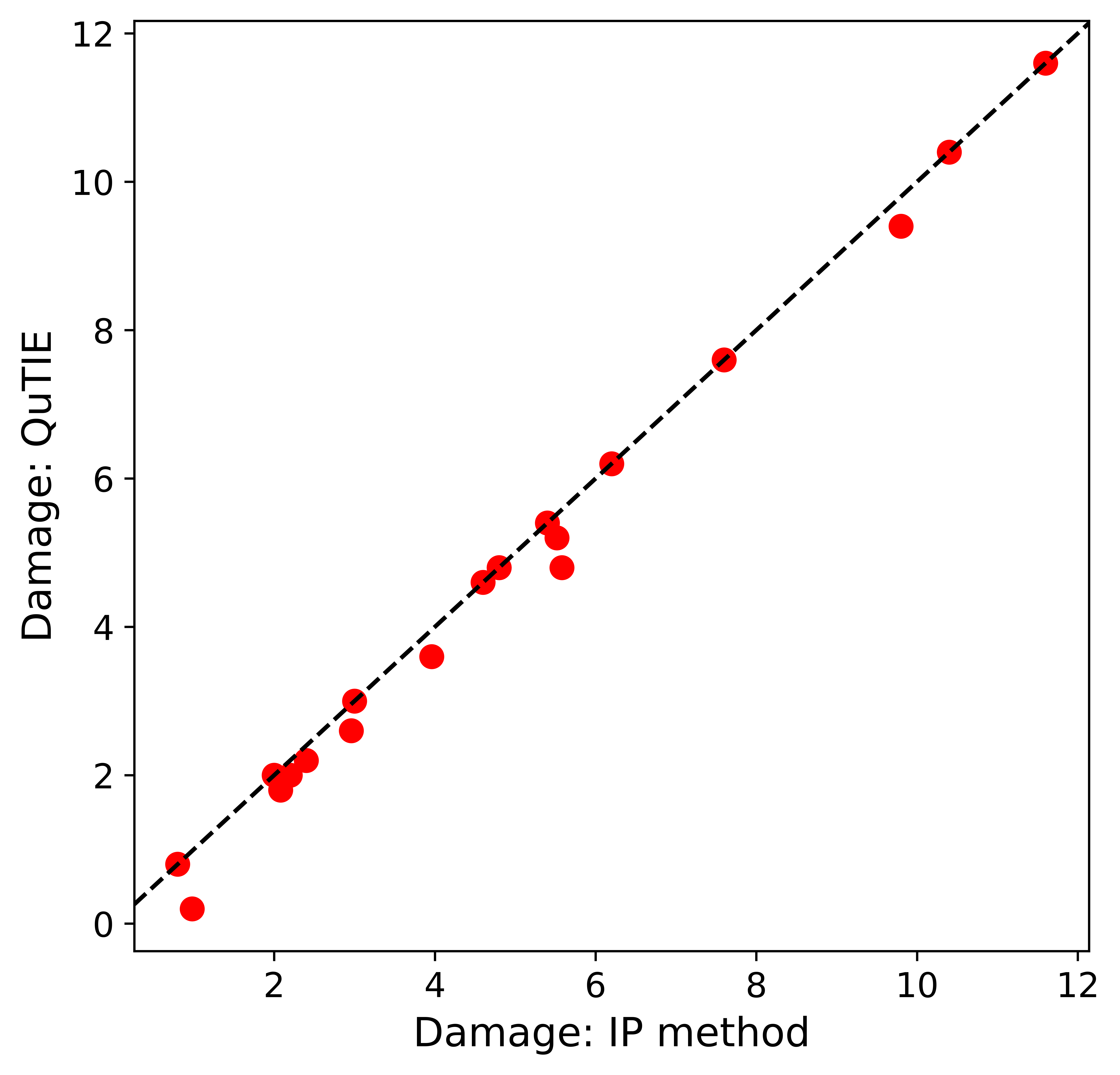}
  \caption{{\em M.Musculus}}
  \label{fig:sfig1_add:mmu}
\end{subfigure}

\caption{Analysis of the IP method and \OurAlgo. (a),(b),(c) Damage values for the three species. Each point corresponds to the average of a combination of one network and one $k$ value across all test cases. The diagonal line is the $x = y$ line
} 
\label{fig:IP:comparison}
\end{figure}

\subsubsection{Comparison with integer programming solution} \label{sec:expt:ip}
Here, we examine the performance of \OurAlgo on small datasets by comparing it against the IP method, which is one of the most popular method for solving optimization problems like TIE~\citep{Tamura2011}.

We compare \OurAlgo and IP in term of damage from resulting solutions. Figure ~\ref{fig:sfig1_add:eco}, \ref{fig:sfig1_add:hsa}, and \ref{fig:sfig1_add:mmu} present the results for three species including E.Coli, H.Sapiens, and M.Musculus respectively. Each point in this figure corresponds to the average damage of one (network, target compound set size) pair. The results demonstrate that \OurAlgo can provide solutions with less damage than IP in most cases (99.3 \% of all test cases). Recall that the IP formulation does not consider the relation of enzymes and reactions at first glance while \OurAlgo, we present the dependency of enzymes and reactions in the Equation~(\ref{eq:main3}). Thus, the results imply that taking the set of enzymes into account is crucial in optimizing damage for TIE problem. 

\subsubsection{Comparison with the double iterative method} \label{sec:expt:heuristic}

Next, we study the performance of \OurAlgo on larger datasets. Exact method does not scale to these networks, so we compare our method to the heuristic Double Iterative method. Our goal in this experiment is to observe whether the damage incurred by the solutions of our method are better than existing heuristic solutions which also scale to large networks and large values of $k$. We use the networks listed in Table~\ref{table:data_large} (see Supplementary Materials 1 - SM 3). In total, we perform 288 experiments (i.e., 4 networks $\times$ 3 species $\times$ 6 values of $k$ $\times$ 2 random repetitions $\times$ 2 methods).

\begin{figure}[t]
    \centering
    \includegraphics[width=0.5\columnwidth]{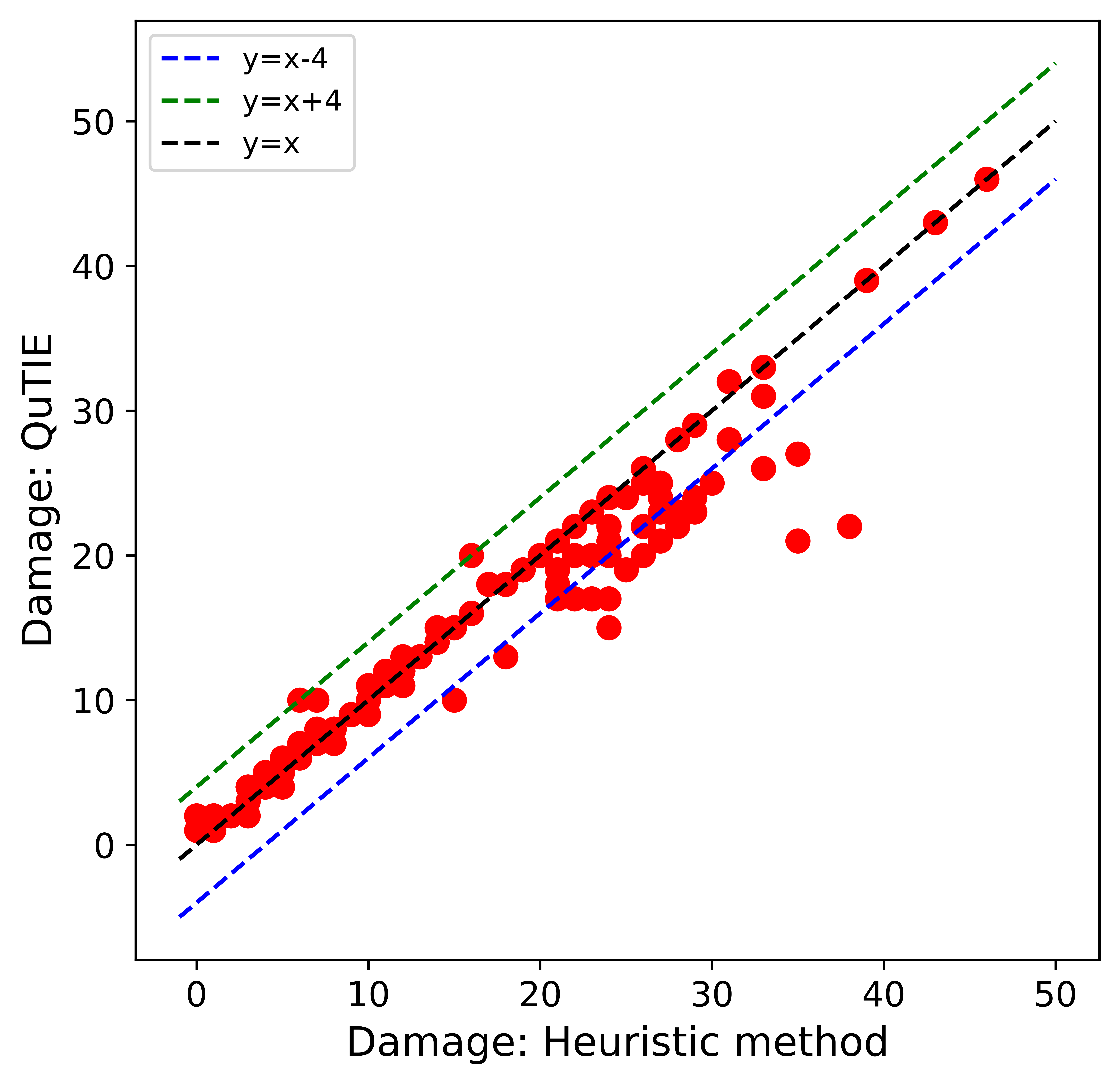}
    \caption{Comparison between \OurAlgo and the heuristic double iterative method in large datasets in term of damage. The less damage is, the better solution is. Data points outside the envelope formed by the green dash line, and the blue dash line indicate cases in which \OurAlgo significantly outperforms the heuristic method.}
    \label{fig:exp3}
\end{figure}

Figure~\ref{fig:exp3} plots the damage values resulting from the two methods for each combination of network and target compound set size.
We observe that \OurAlgo method outperforms the heuristic solution in almost all cases; \OurAlgo identifies a solution with  less than or equal damage than that found by the heuristic method in 117 out of 144 cases (i.e., in 81.2\% of the experiments). In 12.6\% of the experiments, the gap between our method and the heuristic solution is more than 4 in favor of our method, while the heuristic method never yields damage gap $> 4$ in any of the cases. These results suggest that \OurAlgo has potential to identify target enzymes for even large networks when the exact methods do not work without relying on heuristics. The points outside the zone bounded between two lines $y = x + 4$, and $y = x - 4$ are from \emph{Nucleotide metabolism} networks, and \emph{Purine} networks. This suggests that the underlying network topology has great influence on how much \OurAlgo outperforms the competing method.

\ignoreme{
\begin{figure}[h]
    \includegraphics[width=1\columnwidth]{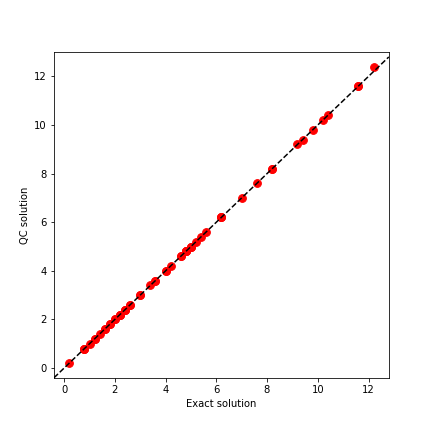}
    \caption{Comparison between QuTIE, and the exact method in small datasets in term of damage. The less damage is, the better solution is. The dash line is the $x=y$ line.}
    \label{fig:exp1}
\end{figure}
}
\ignoreme{
\begin{figure}[h]
    \includegraphics[width=1\columnwidth]{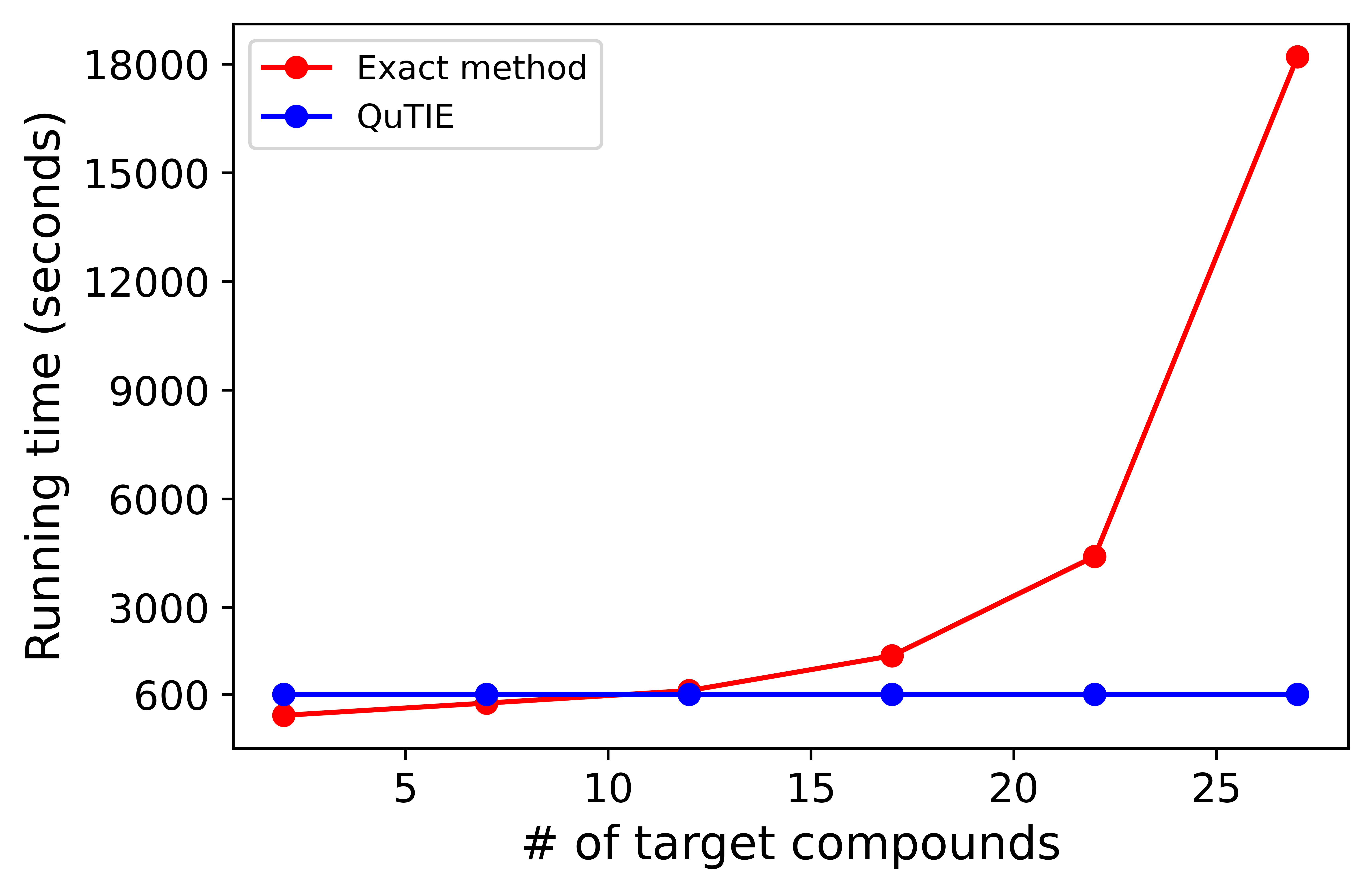}
    \caption{Comparison between \OurAlgo and the exact method in small datasets in term of running time.}
    \label{fig:exp2}
\end{figure}
}


\subsubsection{Comparison with simulated annealing} \label{sec:expt:sa}
\begin{figure}[t]
\centering
    \includegraphics[width=1\columnwidth]{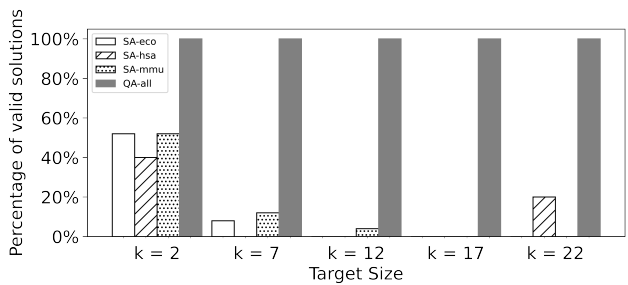}
    \caption{Comparison between \OurAlgo and the SA method on small datasets in term of their success in finding valid solutions.}
    \label{fig:exp_add1}
\end{figure}

Here, we compare the performance of \OurAlgo to its counterpart which is executed on a classical computer on small datasets. Our goal in this experiment is to observe that given a same objective function, whether a quantum computer can explore better solutions through annealing process than a classical computer can. Recall that a solution is valid if it inhibits all target compounds. We compare \OurAlgo and SA in term of the number of times a valid solution can be found over the total number of test cases. We set the time limit for both methods to 10 minutes. The Figure~\ref{fig:exp_add1} shows the percentage of valid solutions for the methods for different datasets and increasing number of target compounds $k$. We observe that \OurAlgo always finds valid solutions. Meanwhile, SA rarely can find valid solutions in cases of $k > 2$ (less than 20 percents). Even in the cases of small target size ($k=2$), SA fails to find valid solutions in 50-60\% of test cases. The results imply that quantum computers can outperform classical computers in solving the TIE problem by simulating annealing process.

\subsection{The impact on actual disease-related compounds}

\begin{figure*}[h]
\begin{subfigure}{.3\textwidth}
  \centering
  \includegraphics[width=1\textwidth]{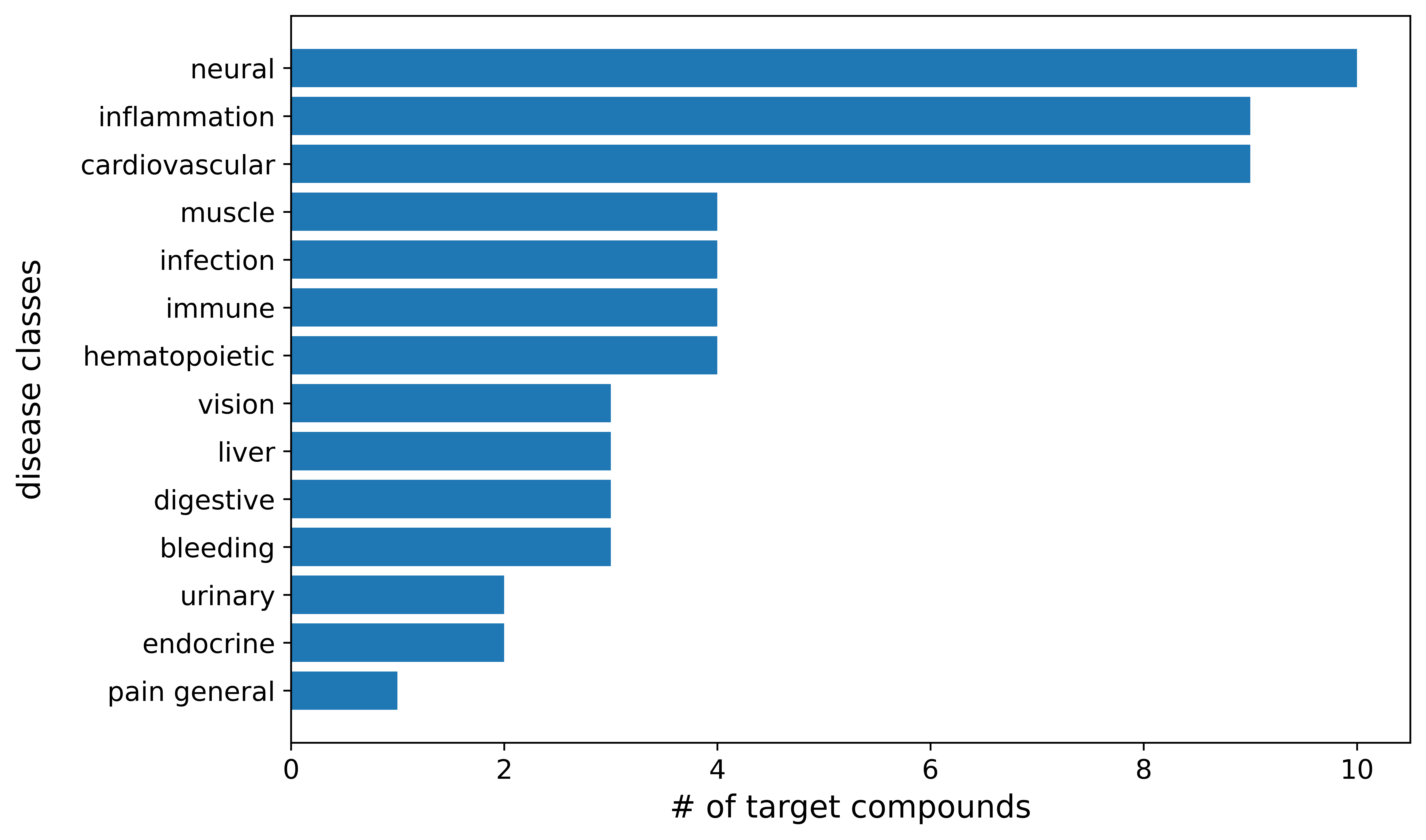}
  \caption{}
  \label{fig:exp41}
\end{subfigure}
\hfill
\begin{subfigure}{.3\textwidth}
  \centering
  \includegraphics[width=1\columnwidth]{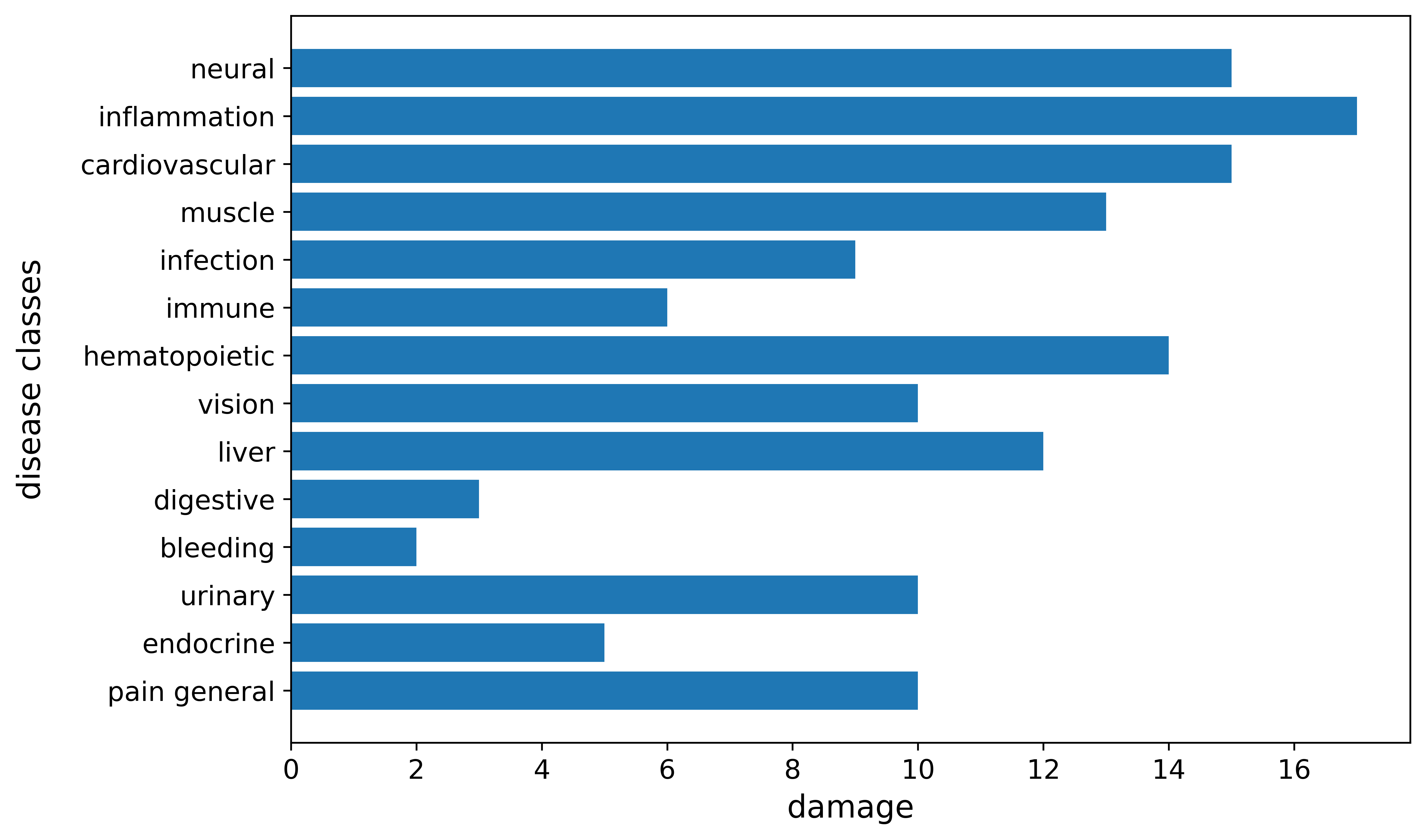}
  \caption{}
  \label{fig:exp42}
\end{subfigure}
\hfill
\begin{subfigure}{.3\textwidth}
  \centering
  \includegraphics[width=1\columnwidth]{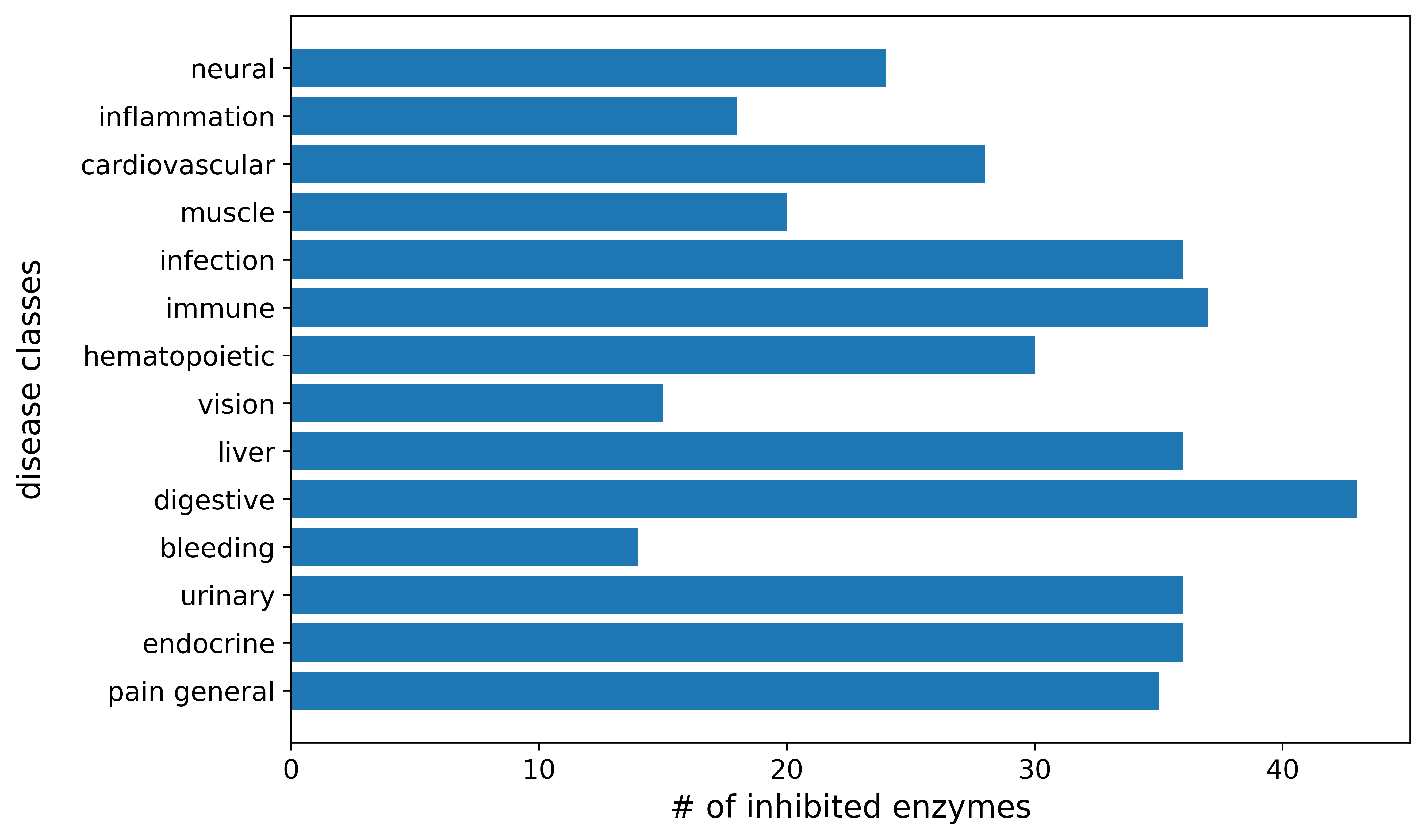}
  \caption{}
  \label{fig:exp43}
\end{subfigure}
    \caption{Analysis of the \OurAlgo in cases of disease-related target compounds in the \emph{Biosynthesis of amino acids} metabolic network. (a) Summary of the number of target compounds related to different disease classes. (b) The correlation between disease classes, and average resulting damage. (c) The correlation between disease classes, and the average number of inhibited enzymes.}
    \label{fig:exp4}
\end{figure*}

So far, we tested our method on real networks, but with randomly selected target compound sets. 
Here, we evaluate how our method performs when the target compound sets are verified to be associated with known disease classes. We use the \emph{Biosynthesis of amino acids} metabolic network of Homo sapiens for every experiments in this part.
We obtain  mapping from disease classes to compounds from the literature
\citep{Zielinski2015}, for 14 major disease classes. Figure~\ref{fig:exp41} shows these disease classes and the number of compounds associated for each disease class in the given metabolic network.
We observe that the number of target compounds shows huge variation among different disease classes (it varies from 1 to 10 with a median of 4). This illustrates the need for new solutions that work well for both small and large target sets.

We run \OurAlgo for each disease class using its corresponding associated compounds as the target compound set. We report the damage as well as the target enzymes identified by our method.
Figure~\ref{fig:exp42} shows the damage \OurAlgo yields for each disease class. Similar to Figure~\ref{fig:exp41}, we observe a huge variation in the damage value (from 2 to 17, with a median of 10). The values in these two figures however are not correlated. That is, smaller target compound set size does not necessarily yield smaller damage. This suggest that the topology of the metabolic network and the distribution of the target compounds over this topology play an important role in the efficacy and the toxicity of the drugs designed for the underlying disease. For example, Figures~\ref{fig:exp41} and~\ref{fig:exp42} together suggest that disease classes bleeding, digestive, and immune deficiencies can be treated with less damage, although they have more target compounds than several other disease classes, such as pain, urinary, and liver failures.

We also examine the number of inhibited enzymes by \OurAlgo for each disease class. This number can be considered as an indicator of the cost/difficulty of inhibiting the enzymes needed to stop the production of those compounds: The larger the number of enzymes, the more effort it takes to inhibit them. Figure~\ref{fig:exp43} presents the results. Similar to Figures~\ref{fig:exp41} and~\ref{fig:exp42}, we observe a huge variation in the number of enzymes (ranging from 14 to over 40). 
It is worth noting that  \OurAlgo aims to minimize the damage, not the number of enzymes. Therefore, it is not surprising that \OurAlgo can sometimes yield a very large number enzymes to obtain smaller damage. We also observe no correlation between the target compound set size and the number of enzymes resulting from them. This also implies that the topology of the distribution of the target compounds on the metabolic network is the primary factor in the size of the resulting enzyme set.

\noindent {\bf Evaluation of disease enzyme associations}
Our final analysis explores whether the targeted enzymes indeed have known associations for the disease classes, whose compounds they inhibit. To do that, we list all the enzymes \OurAlgo identifies as target in order to inhibit the target compounds associated for each disease class. For 14 disease classes, and all the known compounds associated with each disease class, \OurAlgo identifies 408 enzymes as targets in total with repetition (i.e., an enzyme can be a target for multiple disease classes), leading to 53 unique enzymes for all disease classes combined. We provide the list of disease classes, compounds, and enzymes in the Supplementary Materials 2. Further studying these enzymes reveals that \OurAlgo indeed identifies target genes verified to be affecting the target disease in wet-lab experiments on human as well as different animal models. Table~\ref{table:real:evidences} lists five examples out of these due to page limitations. 

The very first target enzyme we identify for the pain category is catechol O-methyltransferase. Studies on both rat and mice models demonstrate that inhibition of catechol O-methyltransferase affects the perception of pain~\citep{KAMBUR2010227}.
Similarly, the first target enzyme we identify for the endocrine disease category is aspartate aminotransferase, whose altered activity in children has a detrimental effect of early-life endocrine-disrupting chemical exposure on liver function~\citep{BARSE200736}.
\OurAlgo identified arginase as one of the top target enzymes for urinary disorders. Indeed, in clinical samples,
of obstructive nephropathy, altered levels of arginase was observed using both Western blot and MRM analysis~\citep{naylor1981}.
Hyperprolinemia is a bleeding disorder, caused by the build up of proline in the blood~\citep{Pandhare2009RegulationAF}. \OurAlgo correctly identifies this enzyme too as a potential drug target.
Finally, the abundance of glutamine, one of the targets we identify for immune related disorders, is indeed linked to the immune supression in humans~\citep{cruzat2018}.
In summary, there is substantial publication evidence supporting the potential target enzymes we identify, which suggest that efficient and accurate solution to the TIE problem on large and complex networks using quantum optimization has great potential to assist drug target identification.

\begin{table}[t]
    \centering
    \begin{adjustbox}{width=0.5\textwidth}
    \begin{tabular}{|l|l|l|}
    \hline
    \textbf{Disease class} &  \textbf{Enzyme name} & \textbf{Evidence}  \\
    \hline
     Pain general &  catechol O-methyltransferase &  \citep{KAMBUR2010227} \\ \hline
     Endocrine  &  aspartate aminotransferase   & \citep{BARSE200736}  \\ \hline
     Urinary  & arginase & \citep{naylor1981}   \\ \hline
     Bleeding  &  prolin oxidase & \citep{Pandhare2009RegulationAF}   \\ \hline
     Immune  & Glutamine synthetase & \citep{cruzat2018}   \\ \hline
    \end{tabular}
    \end{adjustbox}
    \caption{Five example target enzymes identified by \OurAlgo for five disease classes and publication evidences for those enzymes.}
    \label{table:real:evidences}
\end{table}


%
%

\vspace*{-0.5cm}
\bibliographystyle{abbrvnat}
\bibliography{bibliography}

\setcounter{table}{0}
\setcounter{figure}{0} 
\renewcommand{\thetable}{SM. \arabic{table}}
\renewcommand{\thefigure}{SM. \arabic{figure}}

\onecolumn

\pagebreak
\pagestyle{empty}
\section*{{\Large Supplementary Materials 1}}
\section{SM 1: Literature on the TIE problem and limitations}

\begin{figure*}[h]
\begin{subfigure}{.3\textwidth}
  \centering
  \includegraphics[width=1\textwidth]{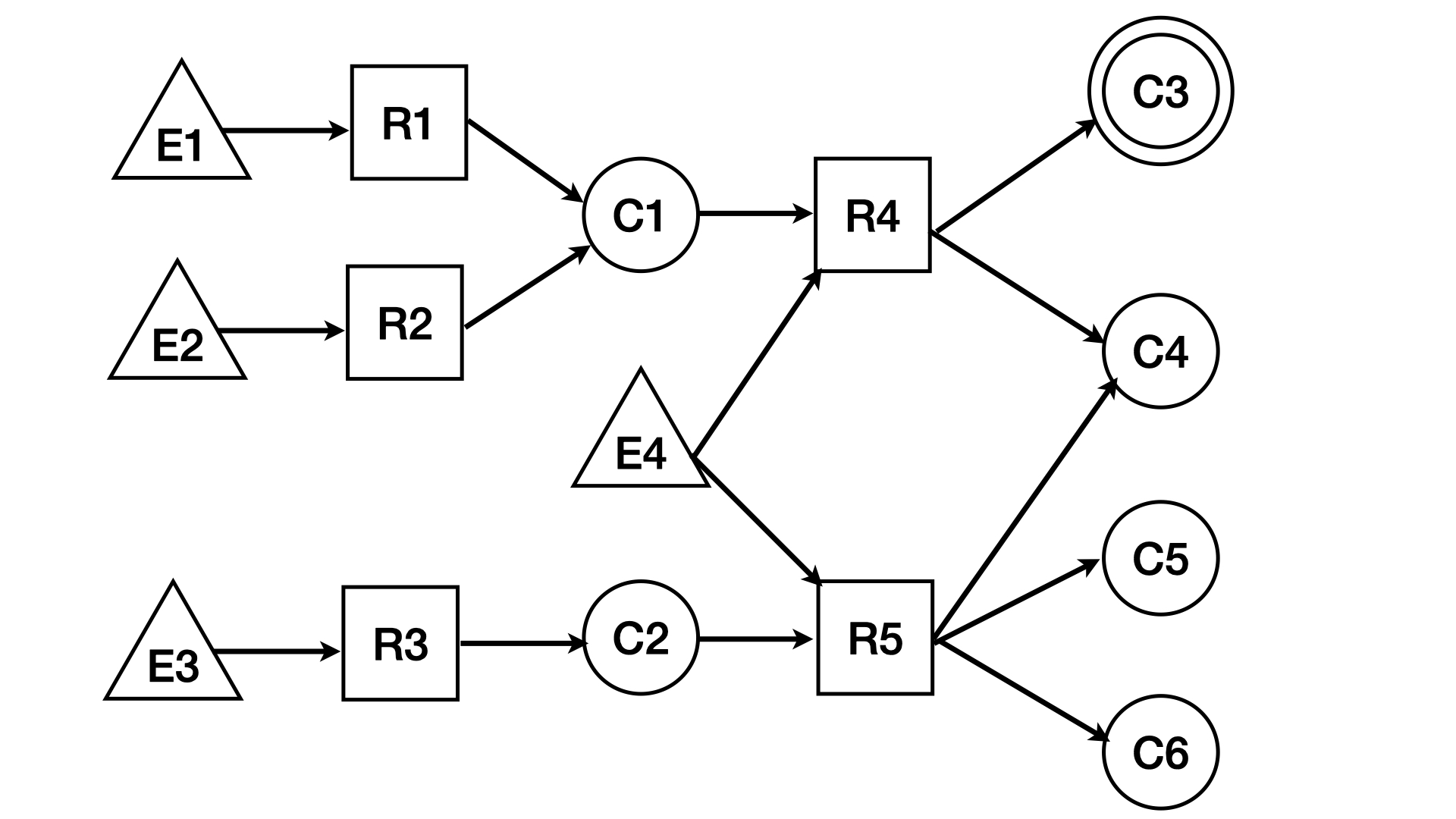}
  \caption{}
  \label{fig:sfig1}
\end{subfigure}
\hfill
\begin{subfigure}{.3\textwidth}
  \centering
  \includegraphics[width=1\columnwidth]{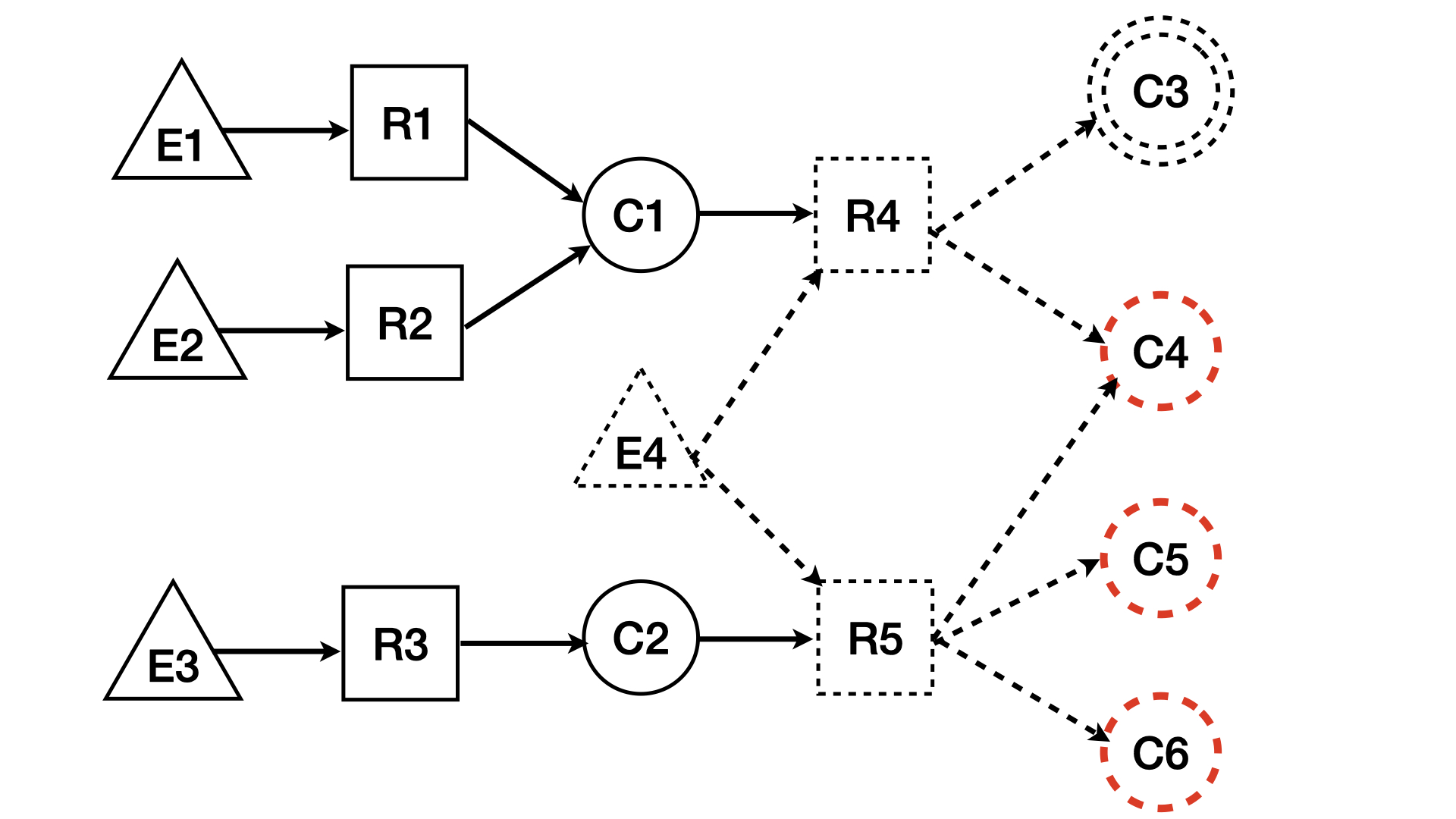}
  \caption{}
  \label{fig:sfig2}
\end{subfigure}
\hfill
\begin{subfigure}{.3\textwidth}
  \centering
  \includegraphics[width=1\columnwidth]{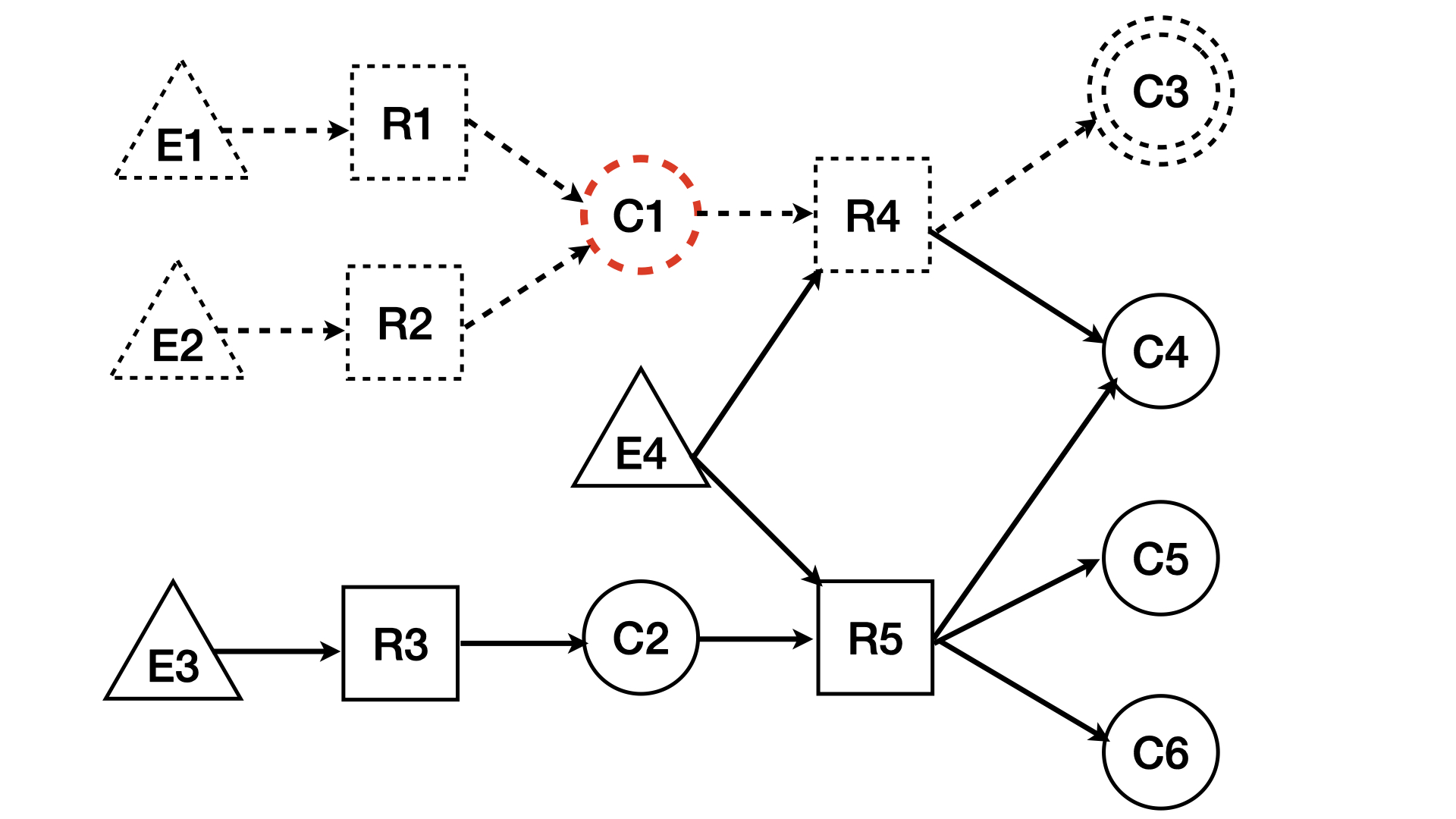}
  \caption{}
  \label{fig:sfig3}
\end{subfigure}

\caption{(a) A metabolic network with a set of enzymes (triangle), reactions (square), and compounds (circle). The target compound is \emph{C3} represented by a double circle. (b) The state of the metabolic network after inhibiting enzyme \emph{E4}. Inhibited nodes are marked by dot-line. Unintentionally inhibited compounds are colored by red. (c) The state of the metabolic network after inhibiting \emph{E1}, and \emph{E2}. }
\label{fig:hypo_img_final}
\end{figure*}

Figure~\ref{fig:hypo_img_final} illustrates an example of the TIE problem. Consider a metabolic network with three enzymes, five reactions and six compounds illustrated in Figure~\ref{fig:sfig1}, our target is to find a set of enzymes such that whose inhibition leads to the elimination of the target compound $C3$ with the least damage. In the Figure~\ref{fig:sfig2}, the inhibition of $E4$ leads to the elimination of $C3$, $C4$, $C5$, and $C6$. Because $C4$, $C5$, and $C6$ are non-target compounds, the damage to the network by inhibiting $E4$ is 3. This selection of inhibited enzymes is not optimal in term of damage. Let's consider another case in which the selection of inhibited enzymes can lead to the optimal damage. If we inhibit the enzymes $E1$ and $E2$, $C1$ is the only non-target compound which is eliminated (see Figure~\ref{fig:sfig3}). Thus, the damage of this enzyme inhibition is 1.

The \emph{OPMET} algorithm solves this problem optimally and improves on the exhaustive search by using branch-and-bound~\citep{Sridhar2008}. However, in the worst case, the complexity of OPMET remains exponential in the number of enzymes in the metabolic network. \emph{Double Iterative} method provides a fast heuristic solution in two phases: \emph{Iterative Phase}, and \emph{Fusion Phase}~\citep{Song2009}. Iterative Phase finds solutions by tracing the network upstream starting from the target compounds. This phase provides feasible solutions quickly, however these solutions might not be optimal (i.e., although their inhibition stops the production of all the target compounds, but they may yield large damage). Fusion Phase takes these solutions,  and then extends the set of enzymes inferred from them by including neighboring enzymes. It then exhaustively searches for optimal solutions in the extended set. This method has two drawbacks. First, the running time of the second phase is exponential in the number of enzymes in the extended set. Second, if the solutions resulting from the Iterative Phase are not diverse enough, the refined solution can be trapped in a local optima. \emph{Minimum Reaction Cuts (BN-ReactionCut)} problem is similar to the TIE problem~\citep{Tamura2011}. Here, the objective is to find the optimal set of reactions such that their inhibition eliminates a given set of target compounds. Tamura et. al, developed an integer programming (IP) for this problem. This solution however ignores the fact that an enzyme can catalyse multiple reactions. As a result, its solutions can yield high damage values to the TIE problem.

\section{SM 2: Background on Quantum Annealing}
In this section, we present the fundamental concepts and terminology needed to understand the Quantum Annealing process and how we solve the TIE problem using QA. We first describe the QA workflow for generalized optimization problems. Next, we describe QUBO representation of optimization problems which can be encoded in QA.

\subsection{Quantum annealing and its work flow}

Quantum annealing (QA) is an extension of simulated annealing - a classical optimization method which is effective for solving multi-dimensional optimization problems with large number of local extrema. QA works following from the principle that a physical system can be described in term of its energy  via a mathematical description, called \emph{Hamiltonian} function. The Hamiltonian function is a mapping from a state of the system to energy. In a physical system, the state in which the Hamiltonian function returns the lowest energy is called the \emph{ground state} of the system. In QA, we define two quantum systems, described by two different Hamiltonian functions, called \emph{initial Hamiltonian}, and \emph{final Hamiltonian}. Initial Hamiltonian is a function whose ground state happens when all qubits are in a superposition state of 0 and 1. The final Hamiltonian represents the optimization problem we want to solve by QA. QA changes the quantum system described by the initial Hamiltonian, into the quantum system described by the final Hamiltonian. As the quantum systems change, they remain in their ground states. Because the system described by the final Hamiltonian represents the optimization problem, the ground state at the end is interpreted as the optimal solution of the optimization problem. Thus, the key challenge in QA is to design the final Hamiltonian.

To construct a final Hamiltonian from an optimization problem, and to encode it to Quantum Processing Unit (QPU), the first step is to define the optimization problem in a special form, called \emph{Ising model}, or its equivalent representation, called the QUBO form. We describe this model and form later. The resulting QUBO formulates the final Hamiltonian, which the quantum system evolves to from an initial Hamiltonian. We encode the resulting QUBO with a graph called logical graph, and embed the logical graph into a QPU whose topology is called hardware graph. 
We refer the interested reader to~\citep{Choi2008} for detailed description of how logical graph is mapped to hardware graph for QA.
To ensure that the optimal solution of QUBO encoded by the resulting embedding is identical to that encoded by the logical graph, we assign appropriate weights for nodes and edges included in the embedding. Finally, after running the annealing process repeatedly on the QPU, we decode the resulting ground state of the final Hamiltonian with the lowest energy to obtain the final solution for the optimization problem.

\subsection{Ising model, and QUBO formulation}
\label{sec:prelim:formulation}

The Ising model uses variables, which take discrete variables modeling dipole magnet of spin states with values in $\{-1, 1\}$ as the basis. 
It expresses a known NP complete problem, called Ising spin glasses problem in decision form. Given $N$ spins with external magnetic field of the $i^{th}$ spins denoted by $h_i$, and coupling strength between the $i^{th}$ and $j^{th}$ spins denoted by $J_{ij}$, the objective is to decide the state of the $i^{th}$ spin denoted by variables $s_i \in \{-1, 1\}$, such that total energy of the system is minimized. Based on above definition of Ising, for $\mathbf{s} = [s_0, s_1, ..., s_{N}]$, we express the given optimization problem in Ising form as follows: 
\begin{equation} \label{eq:ising}
    H_{\textrm{Ising}}(\mathbf{s}) = -\sum_{i=1}^{N} h_is_i - \sum_{i=1}^{N}\sum_{j=i+1}^{N} J_{i,j}s_is_j
\end{equation}

The Ising model deals with inputs belonging $\{-1, 1\}$ basis. By setting $s_i = 2x_i-1$, we obtain an equivalent function $f(x)$ which works on binary basis. In detail, we have:
\begin{equation} \label{eq:ising_to_qubo}
    f(\mathbf{x}) = -\sum_{i=1}^{N} (a_i-2b_i)x_i - \sum_{i=1}^{N}\sum_{j=i+1}^{N} 4J_{i,j}x_ix_j
\end{equation}
with the constants $a_i = \sum^{N}_{j=i+1} 2J_{i,j} +2h_i$, $b_i = \sum^{N}_{j=i+1} J_{j,i}$. In addition, because binary variable $x$ satisfies $x = x^2$, we replace the first term in the Equation~(\ref{eq:ising_to_qubo}) with $\sum_{i=1}^{N} (a_i-2b_i)x_i^2$, and thus transform the Equation~(\ref{eq:ising_to_qubo}) in following form:
\begin{equation} \label{eq:qubo}
    f(\mathbf{x}) = \sum_{i \leq j}^{N} Q_{i,j}x_ix_j
\end{equation}
The function $f$ is called QUBO. Similarly, we convert a QUBO equation back to Ising model. To solve optimization problems by QA, we have to represent our problem in form of Ising with $\{-1, 1\}$ basis or QUBO with binary basis. In this paper, we present our problem in form of QUBO. From Equation~(\ref{eq:qubo}), we express the state with lowest energy of the final Hamiltonian denoted with $\mathbf{x}^{\ast}$ as 
\begin{equation}
    \mathbf{x}^{\ast} = \textrm{argmin}_{\mathbf{x} \in \{0,1\}^N} f(\mathbf{x})
\end{equation}

\subsection{Quantum Hybrid Framework.} 
In order to exploit full potential of quantum computing, as well as available capability of classical computing, we utilize a framework which combines both classical, and quantum computing, called \emph{Quantum Hybrid Framework}. Basically, the hybrid framework includes three main parts: decomposer, sampler, and composer. Decompsers are built based on classical techniques which aims to split the initial problem into smaller parts. For example, one of the techniques used for decompsers, called \emph{Energy-Impact decomposing} breaks down the initial problem into sub-problems such that results from sub-problems contribute energy to the initial problem's result as much as possible. The second part in the framework are samplers, which are pure Quantum Solvers. The role of samplers is to solve resulting sub-problems splitted by decomposers. Because size of sub-problems is much smaller than the initial problem, samplers can provide results efficiently in a small running time. In addition, we can apply the fast Hamiltonian reduction to reduce the number of qubits needed to embed the given problems into QPU~\mbox{\citep{Thai2022}}, facilitating this process. The final part is composers which aim to achieve the final result of the initial problem by combining results of sub-problems. We use Quantum Hybrid Framework provided by D-Wave to solve the QUBO formulation we derive in Section~\ref{sec:qubo:form}.
The framework provides us a solution for our QUBO in a pre-set running time. We then infer the QUBO's solution to a corresponding TIE's solution to evaluate the damage. The Framework also provides us the \emph{total running time} which includes total time needed for the framework to find solutions such as programming time, readout time, anneal time, and delay time. Because the actual running time spent on QPU is not separated from the total running time, we can only estimate how much time quantum computing costs to produce solutions by the total running time.

\newpage
\section{SM 3: Dataset description}

\begin{table}[t] 
    \centering
    \captionsetup{justification=centering}
    \caption{Small networks: Metabolic networks from KEGG database grouped by species. $E$, $R$, $C$, Ed  shows the number of enzymes, reactions, compounds, and edges  respectively. 
    }
    \label{table:data_small}
    \begin{tabular}{ |l|l|l|l|l|l| } 
      \hline
      Species & Functions & $E$ & $R$ & $C$ & Ed \\
     \hline
     &Glycerolipid&14&11&10&37\\
&Citrate Cycle&21&19&20&105\\
{\em E.Coli}&Pentose Phosphate&30&27&23&138\\
&Galactose&30&29&31&149\\
&Glycine,serine, and threonine&35&29&29&122\\ \hline

&Glycerolipid&19&18&13&66\\
&Citrate Cycle&23&21&19&115\\
{\em H.Sapiens}&Pentose Phosphate&24&23&20&119\\
&Galactose&23&21&25&121\\
&Pyruvate&31&25&22&118\\ \hline 
&Glycerolipid&19&18&13&66\\
&Citrate Cycle&23&21&19&115\\
{\em M.Musculus}&Pentose Phosphate&24&23&20&119\\
&Galactose&23&21&25&121\\
&Pyruvate&31&25&22&118\\
     \hline
    \end{tabular}
\end{table}

\begin{table}[t]
    \centering
    \captionsetup{justification=centering}
    \caption{Large networks: Metabolic networks from KEGG database grouped by species. $E$, $R$, $C$, Ed  shows the number of enzymes, reactions, compounds, and edges  respectively.}
    \label{table:data_large}
    \begin{adjustbox}{width=0.5\textwidth}
    \begin{tabular}{ |l|l|l|l|l|l| } 
      \hline
      Species & Functions & E & R & C & Ed \\
     \hline
             &Purine&69&89&66&369\\
{\em E.Coli}&Carbon metabolism&78&65&51&292\\
&Nucleotide metabolism&58&97&55&390\\
&Biosynthesis of amino acids&106&101&98&371\\ \hline
&Purine&75&91&62&391\\
{\em H.Sapiens}&Carbon metabolism&71&59&51&246\\
&Nucleotide metabolism&65&96&53&403\\
&Biosynthesis of amino acids&51&46&49&193\\ \hline
&Purine&77&93&64&398\\
{\em M.Musculus}&Carbon metabolism&71&59&51&246\\
&Nucleotide metabolism&65&96&53&403\\
&Biosynthesis of amino acids&53&47&50&200\\
     \ignoreme{
     eco\_L0&&Purine&69&89&66&369\\
eco\_L1&{\em E.Coli}&Carbon metabolism&78&65&51&292\\
eco\_L2&&Nucleotide metabolism&58&97&55&390\\
eco\_L3&&Biosynthesis of amino acids&106&101&98&371\\ \hline
hsa\_L0&&Purine&75&91&62&391\\
hsa\_L1&{\em H.Sapiens}&Carbon metabolism&71&59&51&246\\
hsa\_L2&&Nucleotide metabolism&65&96&53&403\\
hsa\_L3&&Biosynthesis of amino acids&51&46&49&193\\ \hline
mmu\_L0&&Purine&77&93&64&398\\
mmu\_L1&{\em M.Musculus}&Carbon metabolism&71&59&51&246\\
mmu\_L2&&Nucleotide metabolism&65&96&53&403\\
mmu\_L3&&Biosynthesis of amino acids&53&47&50&200\\
}
     \hline
    \end{tabular}
    \end{adjustbox}
\end{table}
We use metabolic pathways for three species: Escherichia coli (\emph{eco} or {\em E.Coli}), Homo sapiens (\emph{hsa} or {\em H.Sapiens}), and Mus musculus (\emph{mmu} or {\em M.Musculus}) from the KEGG database \citep{Kanehisa2000}. Individual metabolic pathway attaches to a specific function of the species. We categorize these metabolic networks into two groups based on the number of interactions in each: small  and large pathways. 
Tables~\ref{table:data_small} and~\ref{table:data_large} list the networks' characteristics.

\section{SM 4: Correctness of the QuTIE algorithm}

\begin{lemma}
    Let us denote the set of nodes corresponding to each compound in the given metabolic network with $C$, and the inhibition state of each compound $c \in C$ with $x_c$ as explained in Equation~(\ref{eq:indicator:function}). $H_{\text{damage}}$ function in Equation~(\ref{eq:main1}) returns the damage incurred by the states $x_c$ of all the compounds in $C$, multiplied by constant $k_1$.
    \label{lemma:damage}
\end{lemma}

\begin{proof}
From the definition of the variables, $x_c = 1$ only if the compound corresponding to $c$ is inhibited (see Equation~(\ref{eq:indicator:function})). Thus the summation   
$$\sum_{c \in C-C_{\text{target}}} x_c$$ computes the number of inhibited compounds which are not targeted, and thus, it is equal to the damage incurred on the given network. Therefore, Equation~(\ref{eq:main1}) returns the damage incurred by the states $x_c$ of all the compounds in $C$, multiplied by constant $k_1$.
\end{proof}

\begin{lemma}
    Let us denote the set of nodes corresponding to each compound in the given metabolic network with $C$, and the inhibition state of each compound $c \in C$ with $x_c$ as explained in Equation~(\ref{eq:indicator:function}). $H_{\text{target}}$ function in Equation~(\ref{eq:main2}) returns the number of target compounds which are not inhibited by the states $x_c$ of all the compounds in $C$, multiplied by constant $k_2$.
    \label{lemma:target}
\end{lemma}

\begin{proof}
From the definition of the variables, $x_c = 1$ only if the compound corresponding to $c$ is inhibited (see Equation~(\ref{eq:indicator:function})). Each term $(1 - x_c)$ returns 1 if the compound corresponding to $c$ is not inhibited (i.e., $x_c = 0$). It returns 0 otherwise.
Thus the summation   $$\sum_{c \in C_{\text{target}}} (1 - x_c)$$
computes the number of target compounds which fail to be inhibited. Therefore, Equation~(\ref{eq:main2}) returns the number of target compounds which are not inhibited by the states $x_c$ of all the compounds in $C$, multiplied by constant $k_2$.
\end{proof}

\begin{lemma}
    Let us denote the set of nodes in the given metabolic network with $V_G$, and the inhibition state of each entity (compound, reaction, or enzyme) $u \in V_G$ with $x_u$ as explained in Equation~(\ref{eq:indicator:function}). Let us denote the set of nodes corresponding to each reaction with $R$. In addition, we define the set of neighbors of a node $r \in R$ as $N(r)$, and auxiliary binary variables
    $t_{r\alpha}$ with $0 \leq \alpha \leq |N(r)|$ which is equal to 1 if $x_r - \sum_{v \in N(r)} x_{v} + \alpha = 0$, and equal to 0 if otherwise. $H_{\text{reaction}}$ function in Equation~(\ref{eq:main3}) returns 0 only if all the reactions satisfy the inhibition conditions provided in Equation~(\ref{eq:sub1}) by the states as given in $x_u$, $\forall u \in V_G$ and auxiliary binary variables $t_{r\alpha}$ with $r \in R$ and $0 \leq \alpha \leq N(r)$. Otherwise, it returns a positive number.
    \label{lemma:reaction}
\end{lemma}

\begin{proof}
For each node $r \in R$, the inhibition condition provided in Equation~(\ref{eq:sub1}) can be presented by two Inequalities (\ref{eq:sub2}) and (\ref{eq:sub3}). Specifically, the variables $x_r$ with $r \in R$ and $x_v$ with $v \in N(r)$ satisfies Equation~(\ref{eq:sub1}) if and only if they satisfies two Inequalities (\ref{eq:sub2}) and \ref{eq:sub3}.

Let us consider the Inequality~(\ref{eq:sub2}) $$x_r \geq x_v \forall v \in N(r)$$ 
Given that the value of $x_u$ with $u \in V_G$ only receives the value of 0 and 1, Inequality~(\ref{eq:sub2}) holds if and only if $x_r - x_v = 0$ or $x_r - x_v - 1 = 0$ which is equivalent to $(x_r - x_v)^2 + (x_r - x_v - 1)^2 = 1$ for all $r \in R$. In general, Inequality~(\ref{eq:sub2}) holds if and only if Expression 
\begin{equation}\label{eq:suppl:reaction:express:1}
    \sum_{v \in N(r)}[(x_r - x_{v})^2 + (x_r - x_{v} - 1)^2] - |N(r)|
\end{equation} returns the value of 0. Otherwise, the above Expression returns a positive number \textbf{(a)}.

Next, let us consider the Inequality~(\ref{eq:sub3}) $$0 \geq x_r - \sum_{v \in N(r)} x_{v}$$

We denote the difference $x_r - \sum_{v \in N(r)} x_{v}$ as $d_r$ for all $r \in R$. We observe that Inequality~(\ref{eq:sub3}) holds if and only if $d_r \leq 0$. Given that the value of $x_u$ with $u \in V_G$ only receives the value of 0 and 1, the valid values of $d_r$ which satisfy Inequality~(\ref{eq:sub3}) are from $-|N(r)|$ to 0. We have that the auxiliary variable $t_{r\alpha} = 1$ if and only if $d_r + \alpha = 0$ for all $r \in R$. It is equivalent to two Equations as follows:
\begin{equation}\label{eq:suppl:reaction:equation:1}
    x_r - \sum_{v \in N(r)}x_{v} + \sum_{\alpha=0}^{|N(r)|}(\alpha t_{r\alpha}) = 0 \forall r \in R
\end{equation}

\begin{equation}\label{eq:suppl:reaction:equation:2}
    1 - \sum_{\alpha=0}^{|N(r)|} t_{r\alpha} = 0 \forall r \in R
\end{equation}

As a result, Inequality~(\ref{eq:sub3}) holds if and only if Equations~(\ref{eq:suppl:reaction:equation:1}) and Equation~(\ref{eq:suppl:reaction:equation:2}) hold for all $r \in R$. In general, for each $r\in R$, Inequality~(\ref{eq:sub3}) holds if and only if Expression 
\begin{equation}\label{eq:suppl:reaction:express:2}
    \Biggr[x_r - \sum_{v \in N(r)}x_{v} + \sum_{\alpha=0}^{|N(r)|}(\alpha t_{r\alpha}) \Biggr]^2 + (1-\sum_{\alpha=0}^{|N(r)|}t_{r\alpha})^2
\end{equation} returns 0. Otherwise, the above Expression returns a positive number \textbf{(b)}.

By summing up Expression~(\ref{eq:suppl:reaction:express:1}) and (\ref{eq:suppl:reaction:express:2}) for all $r \in R$, we can obtain $H_{\text{reaction}}$. Based on statements \textbf{(a)} and \textbf{(b)}, we can conclude that $H_{\text{reaction}}$ function in Equation~(\ref{eq:main3}) returns 0 only if all the reactions satisfy the inhibition conditions provided in Equation~(\ref{eq:sub1}) by the states as given in $x_u$, $\forall u \in V_G$ and auxiliary binary variables $t_{r\alpha}$ with $r \in R$ and $0 \leq \alpha \leq N(r)$. Otherwise, it returns a positive number.
\end{proof}

\begin{lemma}
    Let us denote the set of nodes in the given metabolic network with $V_G$, and the inhibition state of each entity (compound, reaction, or enzyme) $u \in V_G$ with $x_u$ as explained in Equation~(\ref{eq:indicator:function}). Let us denote the set of nodes corresponding to each compound with $C$. In addition, we define the set of neighbors of a node $c \in C$ as $N(c)$, and auxiliary binary variables
    $w_{c\beta}$ with $0 \leq \beta \leq |N(c)|$ which which is equal to 1 if $x_c - \sum_{v \in N(c)}x_{v} - 1 + \beta = 0$, and equal to 0 if otherwise. $H_{\text{compound}}$  function in Equation~(\ref{eq:main4}) returns 0 only if all the compounds satisfy the inhibition conditions provided in Equation~(\ref{eq:sub6}) by the states as given in $x_u$, $\forall u \in V_G$ and auxiliary binary variables $w_{c\beta}$ with $c \in C$ and $0 \leq \beta \leq N(c)$. Otherwise, it returns a positive number.
    \label{lemma:compound}
\end{lemma}

\begin{proof}
For each node $c \in C$, the inhibition condition provided in Equation~(\ref{eq:sub6}) can be presented by two Inequalities (\ref{eq:sub7}) and (\ref{eq:sub8}). Specifically, the variables $x_c$ with $c \in C$ and $x_v$ with $v \in N(c)$ satisfies Equation~(\ref{eq:sub6}) if and only if they satisfies two Inequalities (\ref{eq:sub7}) and (\ref{eq:sub8}).

Let us consider the Inequality~(\ref{eq:sub7}) $$x_c \leq x_{v}\forall v \in N(c)$$
Given that the value of $x_u$ with $u \in V_G$ only receives the value of 0 and 1, Inequality~(\ref{eq:sub7}) holds if and only if $x_c - x_v = 0$ or $x_c - x_v + 1 = 0$ which is equivalent to $(x_c - x_v)^2 + (x_c - x_v + 1)^2 = 1$ for all $c \in C$. In general, Inequality~(\ref{eq:sub7}) holds if and only if Expression 
\begin{equation}\label{eq:suppl:compound:express:1}
    \sum_{v \in N(c)}[(x_c - x_{v})^2 + (x_c - x_{v} + 1)^2] - |N(c)|
\end{equation} returns the value of 0. Otherwise, the above Expression returns a positive number \textbf{(a)}.

Next, let us consider the Inequality~(\ref{eq:sub8}) $$-|N(c)| \leq x_c - \sum_{v \in N(c)}x_{v} - 1$$

We denote the difference $x_c - \sum_{v \in N(c)} x_{v} - 1$ as $d_c$ for all $c \in C$. We observe that Inequality~(\ref{eq:sub8}) holds if and only if $-|N(c)| \leq d_c$. Given that the value of $x_u$ with $u \in V_G$ only receives the value of 0 and 1, the valid values of $d_c$ which satisfy Inequality~(\ref{eq:sub8}) are from $-|N(r)|$ to 0. We have that the auxiliary variable $w_{c\beta} = 1$ if and only if $d_c + \beta = 0$ for all $c \in C$. It is equivalent to two Equations as follows:
\begin{equation}\label{eq:suppl:compound:equation:1}
    x_c - \sum_{v \in N(c)}x_{v} - 1 + \sum_{\beta=0}^{|N(r)|}(\beta w_{c\beta}) = 0 \forall c \in C
\end{equation}

\begin{equation}\label{eq:suppl:compound:equation:2}
    1 - \sum_{\beta=0}^{|N(c)|} w_{c\beta} = 0 \forall c \in C
\end{equation}

As a result, Inequality~(\ref{eq:sub8}) holds if and only if Equations~(\ref{eq:suppl:compound:equation:1}) and Equation~(\ref{eq:suppl:compound:equation:2}) hold for all $c \in C$. In general, for each $c\in C$, Inequality~(\ref{eq:sub8}) holds if and only if Expression 
\begin{equation}\label{eq:suppl:compound:express:2}
    \Biggl[ x_c - \sum_{v \in N(c)}x_{v}  - 1 + \sum_{\beta=0}^{|N(c)|}(\beta w_{c\beta}) \Biggr]^2 + (1-\sum_{\beta=0}^{|N(c)|}w_{c\beta})^2
\end{equation} returns 0. Otherwise, the above Expression returns a positive number \textbf{(b)}.

By summing up Expression~(\ref{eq:suppl:compound:express:1}) and (\ref{eq:suppl:compound:express:2}) for all $c \in C$, we can obtain $H_{\text{compound}}$. Based on statements \textbf{(a)} and \textbf{(b)}, we can conclude that $H_{\text{compound}}$ function in Equation~(\ref{eq:main3}) returns 0 only if all the compounds satisfy the inhibition conditions provided in Equation~(\ref{eq:sub6}) by the states as given in $x_u$, $\forall u \in V_G$ and auxiliary binary variables $w_{c\beta}$ with $c \in C$ and $0 \leq \beta \leq N(c)$. Otherwise, it returns a positive number.
\end{proof}

\begin{thm}
    Let us denote the set of nodes in the given metabolic network with $V_G$, the inhibition state of each entity (compound, reaction, or enzyme) $u \in V_G$ with $x_u$ as explained in Equation~(\ref{eq:indicator:function}), auxiliary variables $t_{r\alpha}$ with $r \in R, 0 \leq \alpha \leq |N(r)|$, and auxiliary variables $w_{c\beta}$ with $c \in C, 0 \leq \beta \leq |N(c)|$. The values of all the variables $x_u$, $t_{r\alpha}$ and $w_{c\beta}$ which minimize the objective function in Equation~(\ref{eq:objective}) optimally solves the TIE problem.
    \label{theorem:objective}
\end{thm}

\begin{proof}
In  order to prove this theorem, we need to prove two statements: The values of all the variables $x_u$ which minimize the objective function in Equation~(\ref{eq:objective}) satisfy both of the following:
\begin{enumerate}
    \item they are a valid solution (i.e., the given states inhibit all targeted compounds.)
    \item they minimize damage.
\end{enumerate}
The objective function $H$ given in Equation~(\ref{eq:objective}) is the summation of $H_{\text{damage}}$, $H_{\text{target}}$, $H_{\text{reaction}}$ and $H_{\text{compound}}$. From Lemma~\ref{lemma:damage}, \ref{lemma:reaction} and \ref{lemma:compound}, $H_{\text{target}}$, $H_{\text{reaction}}$ and $H_{\text{compound}}$ returns the minimum value of 0 if and only if values of all variables $x_u$, $t_{r\alpha}$ and $w_{c\beta}$ satisfy target inhibition constraint given in Equation~(\ref{eq:sub_add1}) and inhibition conditions given in Equation~(\ref{eq:sub1}) and (\ref{eq:sub6}). Thus, the values of those variables that make the summation of $H_{\text{target}}$, $H_{\text{reaction}}$ and $H_{\text{compound}}$ equal to 0 are a valid solution \textbf{(a)}.

On the other hand, $H_{\text{damage}}$ specifies the damage inferred from the values of those variables. If we set the values of constants $k_2$, $k_3$, and $k_4$ in $H_{\text{target}}$, $H_{\text{reaction}}$ and $H_{\text{compound}}$ respectively big enough such that every positive values of $H_{\text{target}}$, $H_{\text{reaction}}$ and $H_{\text{compound}}$ are bigger than the maximum value of $H_{\text{damage}}$, the value of $H$ for invalid values of variables $x_u$, $t_{r\alpha}$ and $w_{c\beta}$ is always bigger than that for valid values of those variables. In addition, because the values of $H_{\text{target}}$, $H_{\text{reaction}}$ and $H_{\text{compound}}$ for valid values of those variables are all 0, the value of $H$ for valid values of those variables is equal to $H_{\text{damage}}$. Therefore, the values of variables $x_u$, $t_{r\alpha}$ and $w_{c\beta}$ that minimize the objective function $H$ also minimize the damage \textbf{(b)}.

From \textbf{(a)} and \textbf{(b)}, we prove that the values of all the variables $x_u$, $t_{r\alpha}$ and $w_{c\beta}$ which minimize the objective function in Equation~(\ref{eq:objective}) optimally solves the TIE problem.
\end{proof}

\end{document}